\documentclass[11pt,a4paper,aip,jcp]{revtex4-1}

\usepackage{jpcrd2}
\usepackage[latin2]{inputenc}
\usepackage[english]{babel}
\usepackage{booktabs,array,dcolumn}
\usepackage{amsfonts}
\usepackage{graphicx}
\usepackage{color}
\usepackage{amsmath,amssymb}
\usepackage{url}
\usepackage{natbib}
\usepackage{longtable}

\newcommand{\be}{\begin{equation}}
\newcommand{\ee}{\end{equation}}

\newcommand{\cm}{cm$^{-1}$}

\newcommand*{\citen}{}
\DeclareRobustCommand*{\citen}[1]{%
  \begingroup
    \romannumeral-`\x 
    \setcitestyle{numbers}%
    \cite{#1}%
  \endgroup
}

\begin{document}

\title{Definitive ideal-gas thermochemical functions of the H$_2$$^{16}$O molecule}

\author{Tibor Furtenbacher}
\affiliation{
MTA-ELTE Complex Chemical Systems Research Group,
H-1117 Budapest, P\'azm\'any P\'eter s\'et\'any 1/A, Hungary}

\author{Tam\'as Szidarovszky}
\affiliation{
MTA-ELTE Complex Chemical Systems Research Group,
H-1117 Budapest, P\'azm\'any P\'eter s\'et\'any 1/A, Hungary}

\author{Jan Hruby}
\affiliation{Department of Thermodynamics,
Institute of Thermomechanics of the CAS,Dolejskova 5, Prague 8, 
CZ-18200, Czech Republic}

\author{Aleksandra A. Kyuberis}
\affiliation{Institute of Applied Physics, Russian Academy of Science,
Ulyanov Street 46, Nizhny Novgorod, Russia 603950}

\author{Nikolai F. Zobov}
\affiliation{Institute of Applied Physics, Russian Academy of Science,
Ulyanov Street 46, Nizhny Novgorod, Russia 603950}

\author{Oleg L. Polyansky}
\affiliation{Department of Physics and Astronomy, University College London,
London WC1E 6BT, United Kingdom}

\author{Jonathan Tennyson}
\affiliation{Department of Physics and Astronomy, University College London,
London WC1E 6BT, United Kingdom}

\author{Attila G. Cs\'asz\'ar$^{*,}$}
\affiliation{
MTA-ELTE Complex Chemical Systems Research Group,
H-1117 Budapest, P\'azm\'any P\'eter s\'et\'any 1/A, Hungary}

\renewcommand{\baselinestretch}{0.95}
\begin{abstract}
A much improved temperature-dependent ideal-gas internal partition function, 
$Q_{\rm int}$($T$), of the H$_2$$^{16}$O molecule is reported for temperatures 
between 0 and 6000 K.  
Determination of $Q_{\rm int}$($T$) is principally based on the direct summation 
technique involving all accurate experimental energy levels known for H$_2$$^{16}$O 
(almost 20~000 rovibrational energies including an almost complete list 
up to a relative energy of 7500 \cm), 
augmented with a less accurate but complete list of first-principles computed 
rovibrational energy levels up to the first dissociation limit, about 41~000 \cm\  (the 
latter list includes close to one million bound rovibrational energy levels
up to $J = 69$, where $J$ is the rotational quantum number). 
Partition functions are developed for {\it ortho}- and {\it para}-H$_2$$^{16}$O
as well as for their equilibrium mixture.
Unbound rovibrational states above the first dissociation limit are considered 
using an approximate model treatment. 
The effect of the excited electronic states on the thermochemical functions is neglected, 
as their contribution to the thermochemical functions is negligible even at 
the highest temperatures considered.
Based on the high-accuracy $Q_{\rm int}$($T$) and its first two moments, 
definitive results, in 1~K increments, are obtained for the following 
thermochemical functions: 
Gibbs energy, standardized enthalpy, entropy, and isobaric heat capacity.
Reliable approximately two standard deviation uncertainties, as a function of temperature, 
are estimated for each quantity determined.
These uncertainties emphasize that the present results are the most accurate ideal-gas 
thermochemical functions ever produced for H$_2$$^{16}$O.
It is recommended that the new value determined for the 
standard molar enthalpy increment at 298.15~K, 
$9.90404 \pm 0.00001$ kJ~mol$^{-1}$, should replace the old CODATA datum,
$9.905 \pm 0.005$ kJ~mol$^{-1}$.
\\

\noindent Keywords: bound and unbound states; ideal-gas thermochemical quantities;
nuclear motion theory; {\it ortho}- and {\it para}-H$_2$$^{16}$O; partition function; water  

\end{abstract}
\maketitle

\tableofcontents

\listoftables

\listoffigures
 
\newpage
\renewcommand{\baselinestretch}{1.0}
\section{~~Introduction}
Water, the most abundant polyatomic molecule in the universe, plays a major role 
in the radiative balance of the atmospheres of many astronomical objects, including the 
atmosphere of our own earth.\cite{02Bernath,jt437,jt519}
Water is ubiquitous in cool stellar and substellar (brown dwarf) environments 
where it is present over a wide range of temperatures including very high ones 
($T > 3000$ K), outside the range of most experimental laboratory techniques.  
Water is also important for models of combustion systems \cite{94BaCoCoFr,01Burcat}
at medium to high temperatures (though still less than 3000 K).
Predicting high-temperature thermochemical quantities and high-temperature spectra 
of water is important for understanding many of these environments.
The related modeling studies need the accurate knowledge of the partition function,
$Q(T)$, of water from the cold to the hot and some other ideal-gas thermochemical
functions which can be determined straightforwardly \cite{61LeRaPiBr}
from $Q(T)$.

Due to their considerable scientific and engineering interest, 
temperature-dependent thermochemical properties of molecular systems such
as water have been reported in several databases and information 
systems.\cite{01Burcat,82Cox,82McGo,janaf3,89CoWaMe,89GuVeAl,94Pedley,94BeElGu,IAPWS-IF97,janaf4,IAPWS,03FiGaGo.partfunc,jt631,13Ruscic}
Most useful for many practical applications would be real-gas and not 
ideal-gas data,\cite{13Ruscic}
but these are available only for a relatively small number of molecules 
and they are hard to obtain via theoretical (quantum chemical) approaches.  
Ideal-gas data, forming the majority of data in the cited information systems, 
are considerably more straightforward to obtain theoretically.  
As emphasized in standard textbooks,\cite{40MaMa,61LeRaPiBr,00McQuarrie}
all ideal-gas temperature-dependent thermochemical functions can be derived 
from the partition function and its moments.
It would be preferable to obtain accurate temperature-dependent thermochemical 
functions experimentally.  
However, even in the few cases and temperature ranges where this is available 
it is built upon effective (anharmonic) spectroscopic quantities, which 
puts a considerable constraint on the accuracy that can be expected from such studies,
especially at elevated temperatures.  
To accomodate the full temperature range required by the applications, 
one must rely on some sort of 
computation in order to derive the ideal-gas partition and thermochemical functions.

It is important to point out that ideal-gas thermochemistry has been developed
with an emphasis on chemical reactions; thus, only those effects have been considered
important which readily change during a chemical reaction.
A consequence, as noted by Ruscic,\cite{13Ruscic} is that 
``practical {\it thermochemical} functions ignore the overall nuclear spin contribution...
as well as the isotope mixing component, which, in any stoichiometrically balanced
chemical reaction, cancel out across the involved chemical species.''
In spectroscopy, in scientific and engineering applications requiring line-by-line data,
and when treating systems out of equilibrium (see, {\it e.g.}, Ref. \citen{10Krupnov}),
partition functions and thermochemical functions containing nuclear-spin
contributions may be needed.
Thus, in this paper we do consider nuclear spins in our treatment and compute thermochemical
quantities for {\it ortho}- and {\it para}-H$_2$$^{16}$O, as well as for their 
nuclear-spin-equilibrated mixture.
Treatment of the state-independent degeneracy factor is made simple here by the fact that the 
non-permuting $^{16}$O nucleus has zero nuclear spin.

For many semirigid molecules, the simplest analytic technique \cite{00McQuarrie,98Irikura} 
to obtain internal partition functions, namely use
of the harmonic oscillator (HO) and rigid rotor (RR) approximations
for the vibrational and the rotational motions, respectively, yield reasonably 
accurate results at relatively low temperatures (especially around room temperature).  
Partition functions have an integrative nature: 
basically they are a direct sum of weighted energy levels.
This provides much room for approximate treatments; for example,
an approach more sophisticated than the RRHO approximation uses
effective spectroscopic Hamiltonians providing a much improved
estimate for the partition functions and the related thermochemical data, even up to
somewhat elevated temperatures.\cite{97EaRa,97McFl,98AySc}
For water, the perturbative approach
is insufficient and even breaks down at relatively low excitations or,
alternatively, at relatively low temperatures.  
Therefore, to obtain highly accurate, high-temperature partition and thermodynamic
functions for the water isotopologues requires the use of variational
techniques during the computation of the energy levels.\cite{jt309}

A considerable volume of knowledge has been accumulated about computing 
thermochemical functions for H$_2$$^{16}$O.  
Important developments on the computational front include studies by 
Martin {\it et al.},\cite{92MaFrGi_Comp} Harris {\it et al.},\cite{jt228} 
Vidler and Tennyson (VT),\cite{jt263} and others.\cite{13Ruscic,39StMc,93ToZhLiTr,02PrVa}  
Two major sources of high-quality thermochemical data are 
JANAF (Joint Army-Navy-Air Force) \cite{janaf4} and Gurvich,\cite{89GuVeAl,91Gurvich} 
which were originally set up to supply, after appropriate compilation and 
evaluation, thermochemical data for modeling the thermochemistry 
of a large number of small and medium-sized chemical systems.  
For the presentation of the results of this study, the JANAF standard is followed: 
the JANAF-style tables list energy functions, entropies, enthalpies, 
and heat capacities as a function of temperature up to 6000~K.  
The JANAF tables themselves list thermochemical functions from 100~K with 100~K increments, 
but in the present study, due to the high experimental spectroscopic accuracy of our 
lower energy levels, we can list meaningful thermochemical quantities
at even lower temperatures.  
Furthermore, in the Supplementary Material \cite{SupplMat}
to this paper the thermochemical quantities are 
listed at 1~K intervals to ensure that future interpolation efforts could retain the high
accuracy of the present study.
Furthermore, the International Association for the Properties of Water and Steam 
(IAPWS),\cite{IAPWS} an expected user of the data supplied here,
requires thermodynamic data tabulated with this fine granularity.

Our data, with the associated approximately two standard deviation uncertainties, 
should be considered as the most accurate
ideal-gas thermochemical data available for H$_2$$^{16}$O.  
There are several facets of the present study supporting this statement.
Prior to the PoKaZaTeL data \cite{jtpoz} used in the present study ({\it vide infra}),
the most complete {\it ab initio} database for H$_2$$^{16}$O energy levels
and transitions was the so-called BT2 line list,\cite{jt378} which
contains 221~097 energy levels (up to $J = 50$ and $E \le 30~000$ \cm)
and half a billion transitions.
The high-accuracy first-principles PoKaZaTeL dataset employed in this study is complete 
up to the first dissociation limit and contains four times more, 
close to one million energy levels.
Prior to the present work, the most reliable partition sum and related thermochemical 
data was due to VT.\cite{jt263}
VT used a hybrid approach 
similar to the one employed here, but one which was necessarily more approximate.  
They summed over the then available empirical energy
levels,\cite{jt275} augmented with levels from a variational
line list computed by Viti, \cite{jt197,Viti.thesis} and then completed it
with predicted band origins to dissociation \cite{jt230} combined with a 
very approximate treatment of rotation.  
All sums were simply truncated at the
dissociation limit which was assumed to be 41~088 \cm; any states lying
above this limit were ignored.
The present study utilizes a much larger set of experimental energy levels and 
a much larger set of computed first-principles energy levels than 
any of the previous studies.

Due to the Boltzmann distribution characterizing thermodynamic equilibria, 
the contribution of energy levels to the partition function depends strongly 
on the thermodynamic temperature $T$ of the system.  
At the lowest temperatures, where the thermochemical functions depend only on a relatively 
small number of energy levels, an accuracy considerably higher than that provided 
by even the most sophisticated modeling studies can be achieved, once energy levels
of experimental quality are used.  
At the lowest temperatures, one must also be careful how the 
{\it ortho} and {\it para} nuclear-spin isomers of H$_2$$^{16}$O are treated.\cite{10Krupnov}
These isomers are treated explicitly during the present study.

Given the high accuracy we aim at in this study up to very high
temperatures, one must investigate not only the contribution of bound
rovibrational states on the ground electronic state to the
thermochemical functions, but also those of resonance states and 
higher electronic states.  
As shown recently for the case of three
isotopologues of the diatomic molecule MgH,\cite{15SzCsxx.MgH} beyond
a given temperature, dependent upon the first dissociation threshold
of the molecule, unbound states can make a significant contribution to
the partition function and the related thermochemical quantities. 
Studies have begun to consider quasibound states of water,
\cite{jt494,14SzCsxx.H2O} and in this paper such molecular states are
considered for the partition function of water for the first time,
albeit via a very simple model.  

In a complete treatment, the contribution of excited electronic states must also be
investigated.
The effect of the excited electronic states of H$_2$$^{16}$O has not
been considered during the present study, as it deemed to be minuscule even at the
high accuracy sought in this study.

Finally, we note that many of the modeling methods of the present investigation on
H$_2$$^{16}$O can be utilized when determining temperature-dependent thermochemical
functions of other molecular systems.

\section{~~Methodological Details}
The total partition function is assumed to be the product 
of the internal and the translational partition functions.
The bound rovibrational energy levels used for computing the ideal-gas 
internal partition function, $Q_{\rm int}$($T$), of H$_2$$^{16}$O come from two sources: 
a Measured Active Rotational-Vibrational Energy Levels (MARVEL) \cite{07CsCzFuMa,jt412,12FuCs} 
analysis of all the available experimental transitions,\cite{jt539}
and a recent first-principles computation, utilizing the PoKaZaTeL
potential energy surface (PES),\cite{jtpoz} of all the bound rovibrational states 
on the ground electronic state of H$_2$$^{16}$O.  
These two sources will be described separately, followed by a discussion of the 
computation of the thermochemical functions.  
Since it is important to understand the accuracy of all the computed thermochemical quantities, 
an error and uncertainty analysis is also performed as part of this section.

\subsection{~~~~MARVEL energy levels}
The most accurate source of bound rovibrational energy levels of H$_2$$^{16}$O is 
the MARVEL database, obtained as part of an IUPAC-sponsored research 
effort.\cite{jt539,jt454,jt482,jt576,jt562}  
The MARVEL process \cite{jt412} involves a weighted least-squares algorithm, 
whereby first a spectroscopic network (SN) \cite{11CsFu} is built from the experimentally 
observed and assigned (labelled) spectral transitions, involving all available sources of data, 
and then the transitions, based on the Ritz principle, are inverted 
to determine experimental-quality (MARVEL) energy levels.  
Each transition has a label for the upper and lower states between which the transition occurs. 
The labeling scheme for H$_2$$^{16}$O uses six quantum numbers:  
the approximate normal-mode quantum numbers $v_1$, $v_2$, and $v_3$ describe the vibrations 
(symmetric stretch, bend, and antisymmetric stretch, respectively), 
and the usual exact $J$ rotational quantum number and the approximate 
$K_{\rm a}$ and $K_{\rm c}$ values are used for the description of the rotational 
excitation.\cite{Kroto}

The MARVEL database\cite{jt539} for H$_2$$^{16}$O contains 18~486 energy levels, 
all the known and validated experimentally determined bound rotational-vibrational 
energy levels of H$_2$$^{16}$O prior to 2013.  
The uncertainty of the MARVEL energy levels is between $10^{-6}$ and $10^{-2}$ \cm; 
each energy level carries its own uncertainty.
Even with the MARVEL database at hand, complete in rovibrational
energies up to about 7500 \cm, 
one must realize that for higher temperatures (above about 600~K) 
there are insufficient observed rovibrational energy levels available to converge 
the partition function of H$_2$$^{16}$O to 10$^{-4}$ \% accuracy, the
characteristic accuracy below 600 K.  
Therefore, if accurate thermochemical functions are needed at higher temperatures 
one must substantially augment the experimental (MARVEL) set of rovibrational energy levels.  
In the fourth age of quantum chemistry, \cite{12CsFaSzMa} the best way to achieve this 
is through the use of results from first-principles nuclear motion computations,
employing an exact nuclear kinetic energy 
operator and a highly accurate adiabatic global PES.\cite{03PoCsShZo}

\subsection{~~~~First-principles energy levels}
Following this recommendation, in this study the MARVEL energy levels are augmented 
for the bound states by first-principles energy levels.  
The first-principles bound rovibrational energy levels used during this study 
are taken from a database called PoKaZaTeL.\cite{jtpoz}

The PoKaZaTeL energy levels were computed using a global, adiabatic, 
empirically adjusted PES \cite{jtpoz} 
and the DVR3D nuclear-motion code.\cite{jt338}  
This data set contains 810~252 energy levels up to the first dissociation 
limit ($D_0 = 41~145.94(12)$ \cm),\cite{jt549} and it extends all the way to $J = 69$.
As a result, the PoKaZaTeL set represents all the bound rovibrational energy levels
of H$_2$$^{16}$O.

\subsection{~~~~The hybrid database}
The most accurate and most complete database 
of bound rovibrational energy levels of H$_2$$^{16}$O can be obtained 
by combining the complete PoKaZaTeL database with the accurate MARVEL database.
Therefore, we replaced the PoKaZaTeL energy levels with MARVEL energies 
whenever possible and in this way we obtain what is called hereafter the 
hybrid database.

For quantification of the approximately two standard deviation uncertainties 
of the computed thermochemical quantities,
it is essential that each energy level has its own uncertainty. 
The experimental MARVEL energy levels have well determined uncertainties, 
originating from the uncertainties of the measured transitions. 
The computed PoKaZaTeL list does not have associated uncertainties.  
However, by comparing the PoKaZaTeL and MARVEL energy levels, when both are available,
we could estimate the average uncertainties of the PoKaZaTeL energy levels.  
Finally, up to 20 000 \cm\ a value  0.2 \cm\  was taken for these 
one standard deviation uncertainties, 
while above this energy a conservative estimate of 0.5 \cm\  was assumed.

\begin{table}[ht!]
\caption{\label{tab:PhysConst}Physical constants employed in this study.}
        \begin{center}
          \begin{tabular}{llc}
\hline \hline
Name  & Value & Reference \\
		\hline
			Second radiation constant, $c_{2}$~~ &  1.43877736(83) cm K & \citen{CODATA} \\
			Molar gas constant, $R$  &  8.3144598(48) J mol$^{-1}$ K$^{-1}$ & \citen{CODATA} \\
			Avogadro constant, $N_{\rm A}$  &  6.022 140 857(74) $\times$ 10$^{23}$ mol$^{-1}$~~ & \citen{CODATA} \\ 
			Planck constant, $h$  &  6.626070040(81) $\times$ 10$^{-34}$ J s & \citen{CODATA} \\
			Boltzmann constant, $k_{\rm B}$ &1.38064852(79) $\times$ 10$^{-23}$ J K$^{-1}$&\citen{CODATA} \\
			H$_{2}$$^{16}$O molecular mass, $m$~~~ &  2.990724580(36) $\times$ 10$^{-26}$ kg & \citen{ciaaw} \\
\hline \hline
          \end{tabular}
			        \end{center}
\end{table}

\subsection{~~~~Thermochemical quantities}
The internal partition function of a free molecule, $Q_{\rm int}$, 
and its first two moments, $Q_{\rm int}^{'}$ and $Q_{\rm int}^{''}$,
can be written as \cite{61LeRaPiBr,92MaFrGi_Comp,02PrVa}
\be
Q_{\rm int}=\sum\limits_{i} g_i(2J_i+1){\rm exp} \Big(\frac{-c_2E_i}{T}\Big),
\label{Qintnop}
\ee
\be
Q_{\rm int}^{'}=\sum\limits_{i} g_i(2J_i+1) \Big(\frac{c_2E_i}{T}\Big){\rm exp} \Bigg(\frac{-c_2E_i}{T}\Bigg),
\label{Qintp}
\ee
\be
Q_{\rm int}^{''}=\sum\limits_{i} g_i(2J_i+1) \Big(\frac{c_2E_i}{T}\Big)^2{\rm exp} \Bigg(\frac{-c_2E_i}{T}\Bigg),
\label{Qintpp}
\ee
where $c_2=hc/k_{\rm B}$ is the second radiation constant 
(the numerical values of the constants employed in this study are given in Table~1),
$J_i$ is the rotational quantum number, 
$E_i$ is the rotational-vibrational energy level given in \cm, 
$T$ is the thermodynamic temperature in K, 
$g_i$ is the nuclear spin degeneracy factor (representing both state-dependent
and state-independent elements), and the index $i$ runs over all possible 
rovibronic energies considered.  
In the case of H$_2$$^{16}$O, the values of $g_i$ are taken as 3 for the {\it ortho} and 
1 for the {\it para} nuclear-spin states, 
in accord with the HITRAN convention.\cite{03FiGaGo.partfunc}

The full partition function $Q$ of a molecule in the ideal gas state is 
a product of the internal partition function, $Q_{\rm int}$, and the
translational partition function, $Q_{\rm trans}$.
The latter can be expressed as \cite{00McQuarrie}
 \be
    Q_{\rm trans}=V \Lambda^{-3}\,,
 \ee
where $V$ is the volume of the system,
$\Lambda = h/(2\pi m k_{\rm B} T)^{1/2}$ is the de Broglie wavelength, 
$h$ is the Planck constant, and $m$ is the molecular mass 
(the numerical values of the constants are given in Table~1).

The Helmholtz energy $A$, the internal energy minus the product of
thermodynamic temperature and entropy, is obtained from its fundamental relation to the 
canonical partition function $Q$, namely
\be
    A = -RT \ln Q = -RT \ln Q_{\rm int} - RT \ln \frac{V}{\Lambda^3}\,,
\ee
where $R$ denotes the molar gas constant (Table ~1). 
All thermochemical functions can then be derived using thermodynamic identities; 
in particular,
\be
    p=-\frac{\partial A}{\partial V}\,,\quad
    S=-\frac{\partial A}{\partial T}\,, \quad
    G=A+pV\,,\quad
    H=G+TS \,,
\ee
where $p$, $S$, $G$, and $H$ are pressure, entropy, 
Gibbs energy, and enthalpy, respectively. 
The first relation obviously results in the ideal gas equation of state, $pV=RT$. 
The isochoric heat capacity is obtained as
\be
    C_{v}=T\frac{\partial S}{\partial T}=-T\frac{\partial^2 A}{\partial T^2}\,,
\ee
and the isobaric heat capacity of the ideal gas is then $C_{p}=C_{v}+R$.
All these properties can be obtained using the internal partition function, 
Eq.~(1), and its first two moments, Eqs.~(\ref{Qintp}) and (\ref{Qintpp}).
The most important and widely used thermochemical functions can be constructed as follows:

(a)	The standardized enthalpy is
\be
    H(T)-H(298.15)=RT\frac{Q_{\rm int}^{'}}{Q_{\rm int}}+\frac{5}{2}RT-H(298.15),
\label{hcf}
\ee
where $H(298.15)$ is the (absolute) enthalpy at the reference temperature taken to be 298.15~K.

(b)	The Gibbs energy function is
\be
		{\rm gef}(T,p)=-\frac{G(T)-H(298.15)}{T}=R \ln Q_{\rm int}
        +R\ln \frac{(2\pi m)^{3/2}(k_{\rm B}T)^{5/2}}{h^3 p}
        +\frac{H(298.15)}{T}\,.
\label{gef}
\ee

(c)	The entropy is
\be
		S(T,p)=R\frac{Q_{\rm int}^{'}}{Q_{\rm int}}+R{\rm ln}Q_{\rm int}
    +\frac{5}{2}R+R\ln \frac{(2\pi m)^{3/2}(k_{\rm B}T)^{5/2}}{h^3 p}\,.
\label{Entr}
\ee

(d)	The isobaric heat capacity is
\be
		C_{p}(T)=R \left[\frac{Q_{\rm int}^{''}}{Q_{\rm int}}-\left(\frac{Q_{\rm int}^{'}}{Q_{\rm int}}\right)^2 \right]+\frac{5}{2} R.
\label{Cp}
\ee

As seen in Table~\ref{tab:PhysConst}, 
the physical constants used in Eqs. (\ref{Qintnop}) to (\ref{Cp}), 
similarly to the energy levels, have well defined uncertainties.
The uncertainties of the $c_2$ and $R$ constants are rather substantial,
in fact larger than the relative uncertainties of many of the MARVEL energy levels.
Since $c_2$ appears alongside the $E_i$ energies in Eqs. (\ref{Qintnop}) to (\ref{Qintpp}),
its uncertainty has a significant effect on the uncertainties of the computed
thermochemical functions ({\it vide infra}).

\subsection{~~~~The effect of unbound states on the thermochemical properties of water}
A possible route to determine the $Q_{\rm U}$($T$) contribution of the unbound 
rovibrational states to the partition function $Q_{\rm int}$($T$) of water 
is through the use of the expression
\be
		Q_{\rm U}(T)=\int_0^\infty \! \rho_{\rm U}(E){\rm exp}(-\beta E)\mathrm{d}E,
\label{QU}
\ee
where $\rho_{\rm U}$($E$) is the density of the unbound rovibrational states 
for H$_2$$^{16}$O and $\beta = 1/k_{\rm B}T$.  
In the present work, a simple model is used to evaluate Eq.~(\ref{QU}): 
the unbound (scattering/continuum) states of the H$_2$$^{16}$O system are approximated 
as the eigenstates of the non-interacting bound OH radical and an OH + H scattering system, 
in which the OH is treated as a particle with no internal degrees of freedom.  
The density of states for the OH radical can be given by
\be
		\rho^{\rm (OH)}(E)=\sum\limits_{l,v}(2l+1)\delta \Big(E-E_{l,v}^{\rm (OH)}\Big),
\label{rhoOH}
\ee
while for the OH + H scattering system it is

\be
\rho_{\rm U}^{\rm (OH + H)}(E)=\frac{1}{\pi}\sum\limits_j(2j+1)\frac{\mathrm{d}\eta_j(E)}{\mathrm{d}E},
\label{xplus2}
\ee
where $l$ and $v$ are the rotational and vibrational quantum numbers of the OH radical, respectively, 
$j$ is the rotational quantum number of the H + OH scattering system 
in the center-of-mass frame, and $\eta_j(E)$  
is the scattering phase shift corresponding to a given $j$.

Applying the formula, motivated by the probability density distribution formula 
for the sum of two independent random variables,
\be
		\rho_{\rm U}(E)=\int_0^\infty\rho^{\rm (OH)}(E{'})\rho_{\rm U}^{\rm (OH+H)}(E-E{'})\mathrm{d}E{'},
\label{xplus3}
\ee
for the total density of states, and from combining 
Eqs.~(\ref{QU}), (\ref{rhoOH}), (\ref{xplus2}), and (\ref{xplus3}) 
and utilizing the fact that $\eta_j(E)$  is zero for $E < D_0$, one obtains
\be
\begin{aligned}
Q_{\rm U}(T)=
\Bigg(\sum\limits_{l,v}(2l+1){\rm exp} \Big(-\beta E_{l,v}^{\rm (OH)}\Big)\Bigg) \Bigg(\frac{1}{\pi}\sum\limits_j(2j+1)\int_{D_0}^{\infty}\frac{\mathrm{d}\eta_j(E^{'})}{\mathrm{d}E^{'}}{\rm exp} \big(-\beta E^{'}\big)\mathrm{d}E^{'}\Bigg) = \\
Q^{\rm (OH)}(T)Q_{\rm U}^{\rm (OH+H)}(T).
\label{xplus4}
\end{aligned}
\ee
Thus, the total partition function is a product of the partition functions of 
the non-interacting subsystems, as expected.  
The final $Q_{\rm U}$($T$) values were obtained for {\it ortho}- and 
{\it para}-H$_2$$^{16}$O by multiplying 
the results of Eq. (\ref{xplus4}) by 3 and 1, respectively.

The potential energy curves (PEC) for the OH radical and the OH~+~H system 
were obtained from the global H$_2$$^{16}$O PES of Refs.~\citen{03PoCsShZo}
and \citen{06BaShZoPo}.  
The OH PEC was simply obtained by setting the second H to a 30~$a_0$ distance 
from the OH center-of-mass, while for the OH~+~H system the PEC was obtained 
by ``relaxing'' the orientation of the OH and the OH distance, within 0--3~$a_0$, 
for each fixed OH--H distance.  
Eigenenergies for the OH radical in Eq.~(\ref{xplus3}) were obtained by 
solving the diatomic rovibrational time-independent Schr\"odinger equation  
using 250 spherical-oscillator DVR basis functions \cite{10SzCsCz} 
with $R_{\rm max} = 15$~$a_0$.  
As in previous studies,\cite{15SzCsxx.MgH,82MiJu} 
the scattering phase shifts in Eq.~(\ref{xplus2}) 
were computed using a semi-classical WKB approximation.  
The maximum $j$ value used in Eq. (\ref{xplus2}) is 278.

\begin{figure}
\includegraphics[width=0.45\textwidth]{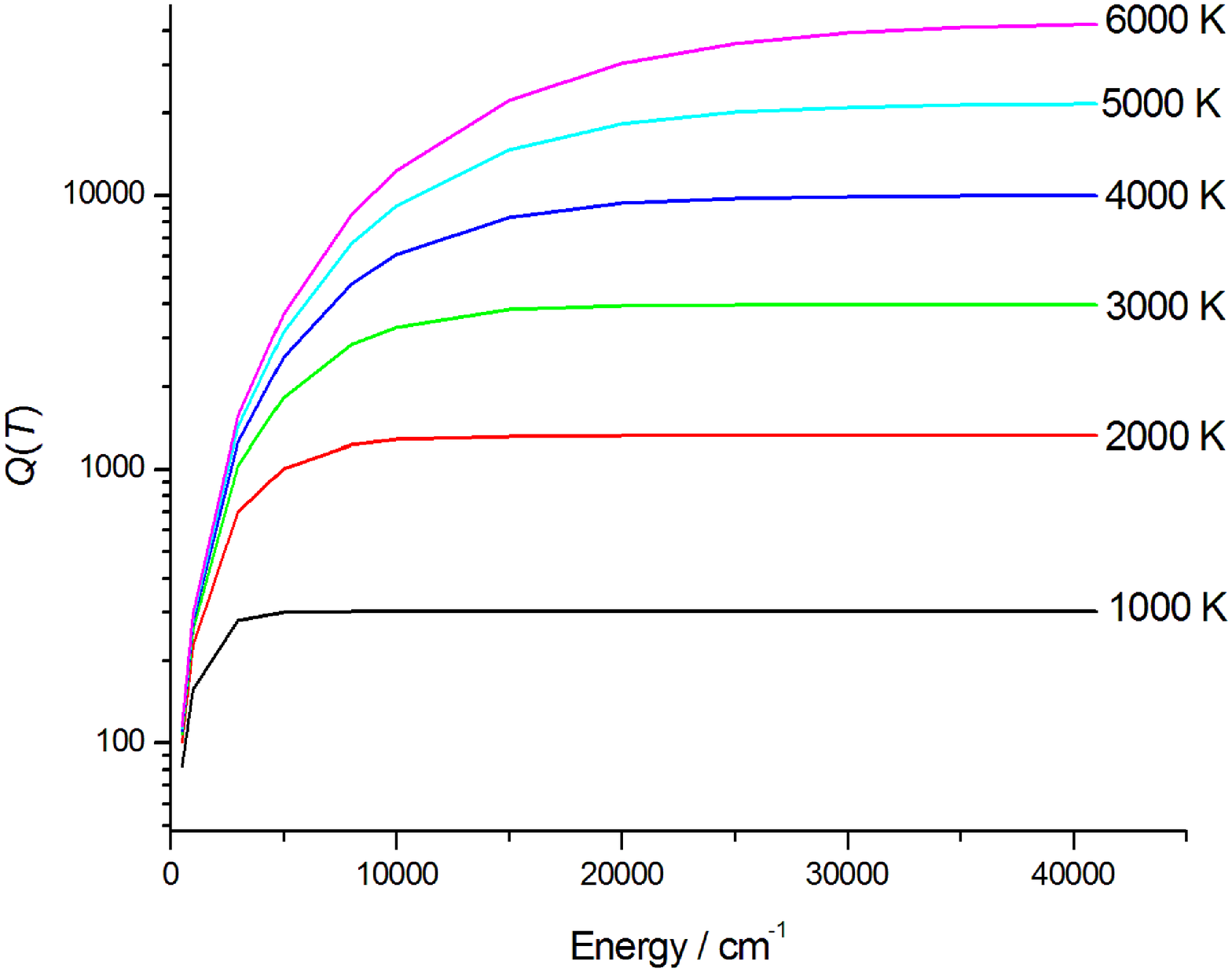}
\includegraphics[width=0.45\textwidth]{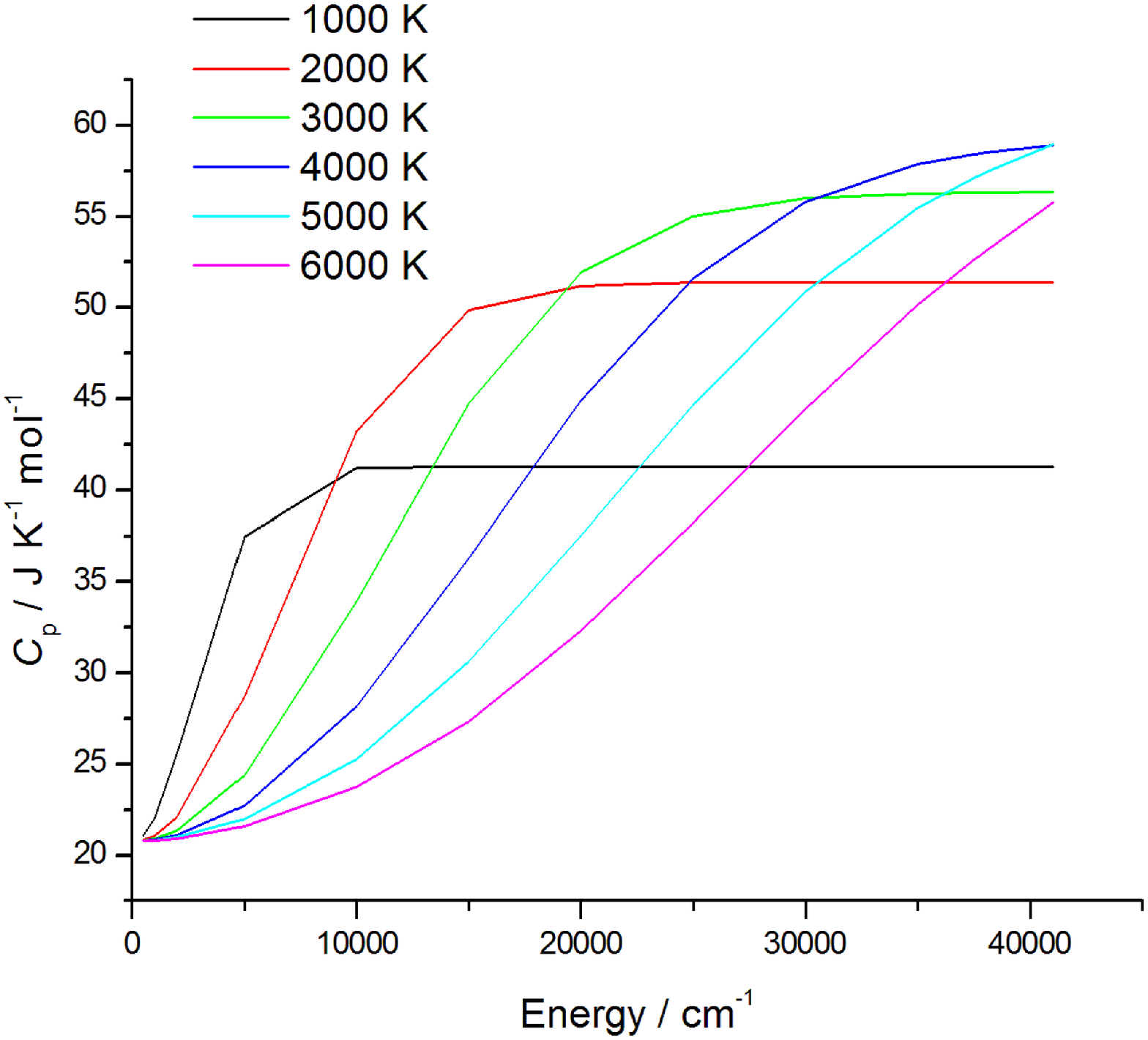}
\caption{Left panel: Convergence of the internal partition function 
$Q_{\rm int}$($T$) of H$_2$$^{16}$O at different temperatures as a function 
of the energy cutoff value considered in the direct sum (see text).
Right panel: Similar curves for the isobaric heat capacity $C_{p}$($T$) of H$_2$$^{16}$O.}
\label{fig:QconvT}
\end{figure}

\subsection{~~~~Uncertainty and error analysis}
The exact values of the internal partition functions of molecules are unknown, and thus 
there are no true reference values available for comparison with the approximate values.
Nevertheless, in a computational study claiming high accuracy a quantification of
uncertainties must be performed.\cite{16ChBrBaCs}

There are a few sources of error preventing the determination of 
``exact'' values of $Q_{\rm int}$($T$).
Traditionally, the largest source of the uncertainty in a partition function, 
especially at higher temperatures, has been the uncertainty about the number 
of bound energy levels (uncertainty about the energy level density). 
A second significant source of error lies in the uncertainty of the energy levels 
used to determine $Q_{\rm int}$($T$).  
A third type of (usually less significant) uncertainty is connected with the question 
of how unbound states and states 
associated with excited electronic states should be accounted for.
A fourth source of uncertainty, so far left unexplored in computational
thermochemical studies, is connected to the uncertainty of the physical constants
entering Eqs. (1) through (16) (Table~1).

Checking the convergence of partition functions is very hard, since $Q_{\rm int}$ grows 
monotonically as more and more bound energy levels are considered in the direct sum.  
At low temperatures ($T < 1000$ K), relatively few energy levels are sufficient 
to reach a converged $Q_{\rm int}$ value 
(in our definition this means that adding more and more higher-lying energy levels 
to the sum in Eq. (1) causes only a negligible change, (much) less than 0.01 \%).

Two simple methods can be used for obtaining the second type of uncertainty mentioned
about the partition function and the associated thermochemical quantities: 
in method A the common error propagation formula can be employed, while 
method B increases and reduces the energy levels by their uncertainties, the two extrema 
of the given thermochemical function can be calculated and the difference of these extrema 
provides an uncertainty estimate.

The third type of uncertainty of $Q_{\rm int}$($T$) comes from the unbound states but,
to the best of our knowledge, this uncertainty has not been taken properly into account
for molecules containing more than two atoms.
Part of the reason is that unbound states start playing a significant role 
at higher temperatures and
only for molecules with a comparatively low dissociation energy.
In this work, the effect of unbound states is approximated using the model described 
in the Sec. 2.5.
In the case of bound states, where very accurate reference data are available, 
the accuracy of the crude ``non-interacting OH plus OH+H'' model for computing 
thermodynamic properties can be tested. 
In fact, this model overestimates the partition function by a factor of around four. 
This huge discrepancy is probably due to the fact that the model allows for quantum states 
with large overlaps between the hydrogen nuclei, which 
in a more realistic simulation would lead to very high (even unbound) energies 
and much smaller contributions to the partition function. 
The situation is expected to be similar in the case of unbound states, that is, 
the model defined in Sec. 2.5 is expected to overestimate the contribution 
of the unbound states in the partition function. 
Thus, taking the computed values of the contribution of unbound states themselves 
as the uncertainties originating from the unbound states seems to be a safe, 
conservative estimate.

In the present case of H$_2$$^{16}$O, 
the hybrid database contains all the existing bound rovibrational energy levels.  
Completeness of the set of hybrid energy levels may not be maintained perfectly 
just slightly below the first dissociation limit, 
where hard-to-determine long-range states may exist;\cite{jt358,10SzCsCz}
therefore, it is worth checking the convergence of $Q_{\rm int}$($T$) by increasing the 
number of energy levels considered in the direct sum via moving an $E_{\rm cut}$ cutoff
energy value closer and closer to the dissociation limit. 
Figure~\ref{fig:QconvT} illustrates the effect of the increase of $E_{\rm cut}$ on the 
total partition sum at different temperatures. 
It can be seen that, while at 1000 K the partition sum is fully converged with an
$E_{\rm cut}$ of about 8000 \cm, 
at 6000 K the $Q_{\rm int}$($T$) does not reach full convergence even at $E_{\rm cut}=D_0$, 
so adding new (high-lying) energy levels to the direct sum the value of the partition 
function might still change noticably.

\begin{figure}
\includegraphics[width=0.45\textwidth]{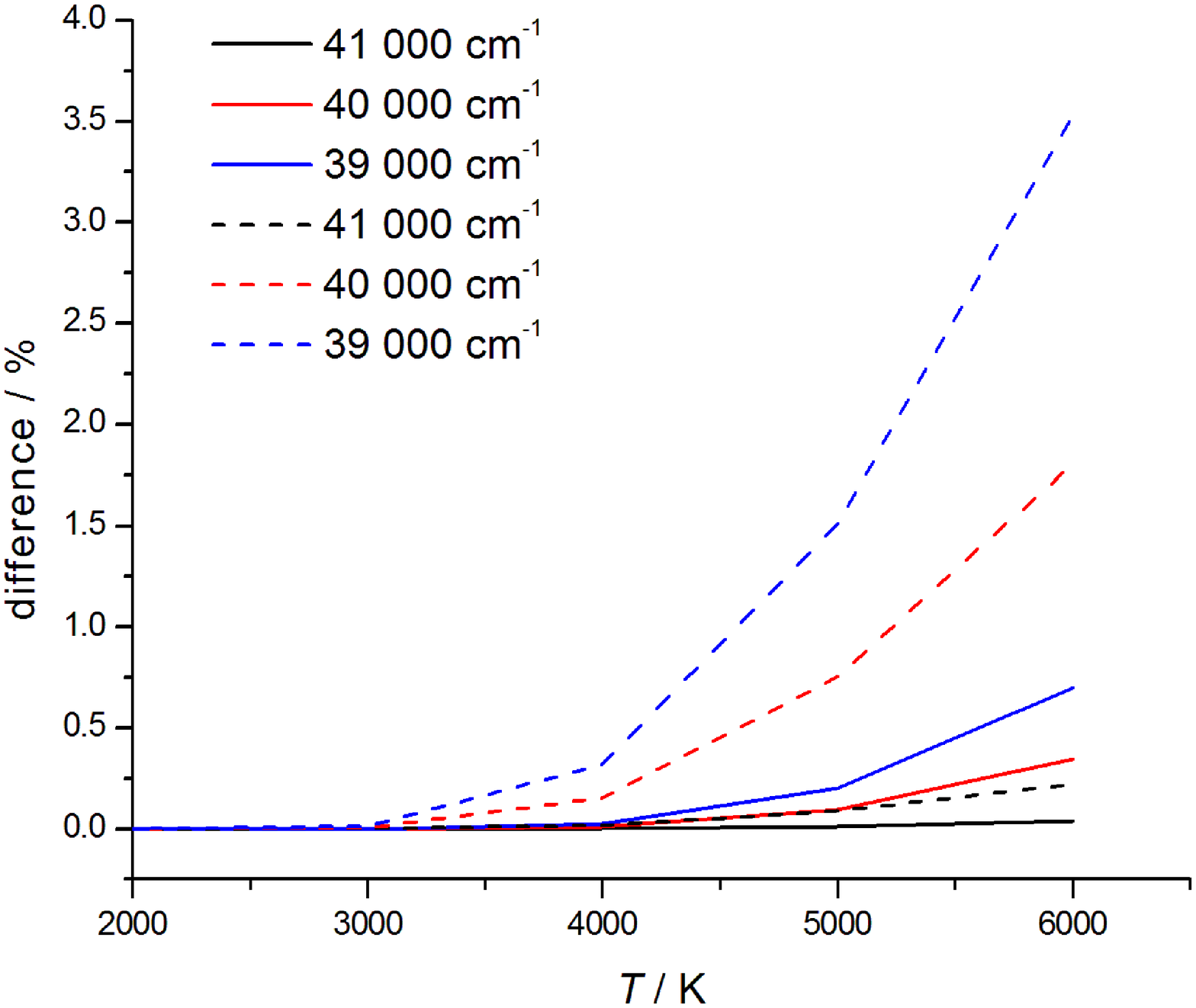}
\includegraphics[width=0.45\textwidth]{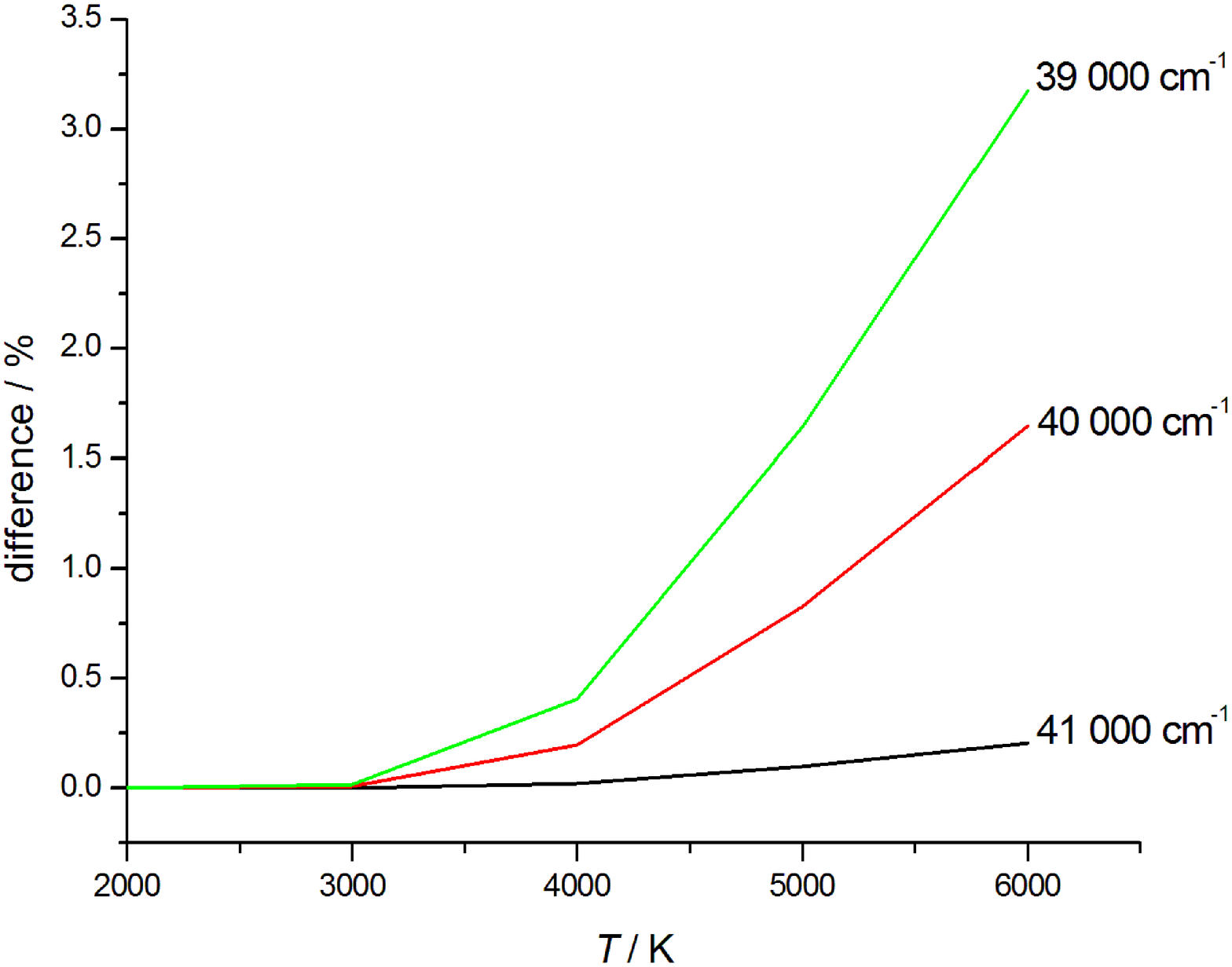}
\caption{Left panel: Convergence characteristics of the 
internal partition function, $Q_{\rm int}$($T$) (solid lines), and the second moment 
of $Q_{\rm int}$($T$), $Q_{\rm int}''$($T$) (dashed lines), of H$_2$$^{16}$O
utilizing larger and larger sets of energy levels, as a function of temperature, 
with energy cutoff values given in the inset of the figure. 
Right panel: Similar curves
for the isobaric heat capacity $C_{p}$($T$) of H$_2$$^{16}$O.}
\label{fig:QconvE}
\end{figure}

To help elucidate the results of Fig.~\ref{fig:QconvT},
the solid lines in Fig.~\ref{fig:QconvE} show the difference, in \%, 
between $Q_{\rm int}^{\rm tot}$ 
(considering all energy levels) and $Q_{\rm int}^{\rm 39000}$, $Q_{\rm int}^{\rm 40000}$, 
and $Q_{\rm int}^{\rm 41000}$ 
(\textit{i.e.}, considering the energy levels up to $E_{\rm cut}=39~000$, 40~000,
and 41~000 \cm, respectively) as a function of temperature. 
It can be seen that 
(a) at 4000 K the differences are still very close to zero;
and (b) at 6000 K the difference between 
$Q_{\rm int}^{\rm tot}$ and $Q_{\rm int}^{\rm 41000}$ is about 0.05\%. 
The dashed lines in the left panel of Fig.~\ref{fig:QconvE} show the 
similar differences for $Q''_{\rm int}$. 
It can be seen that at 6000 K the error of $Q^{'' 41000}_{\rm int}$ is about 0.3\%. 
Considering that there are almost 13~000 energy levels between 41~000 \cm\   and the 
first dissociation limit, and that this number probably grossly overestimates 
the number of energy levels that sophisticated first-principles computations can miss,
we associate the differences of the $Q_{\rm int}^{\rm tot}$ and $Q_{\rm int}^{\rm 41000}$ 
values with the uncertainty which comes from the lack of a truly complete set of bound 
rovibronic energy levels. 
Figure~\ref{fig:QconvE} also shows why it is so important to determine all 
rovibrational energy levels up to the dissociation limit. 
At higher temperatures ($T > 3000$ K), the lack of rovibrational energy levels
at the highest level density regions close to dissociation causes significant errors, 
especially in the cases of $Q^{''}$ and $C_{p}$.

\begin{figure}
\includegraphics[width=0.75\textwidth]{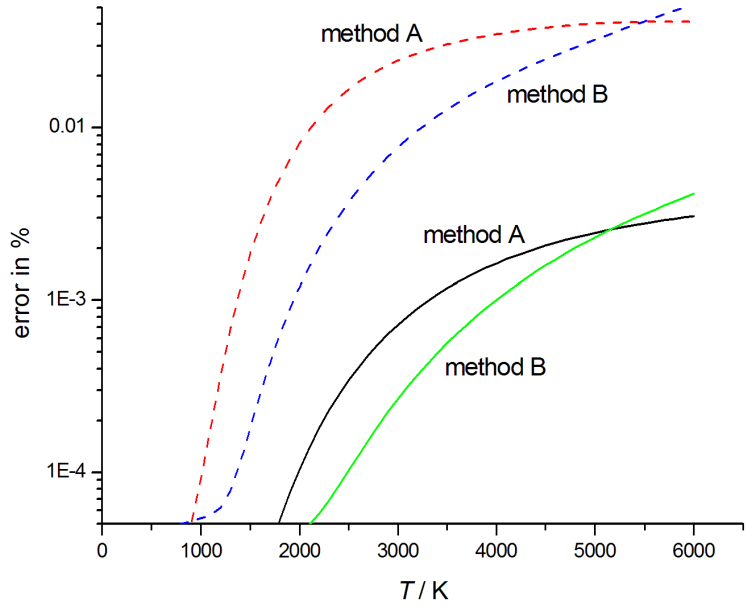}
\caption{Error of the partition function (solid lines) and its second moment (dashed lines) 
using the propagation formula method, Method A, and the ``two extrema'' method, Method B.}
\label{fig:Qerror}
\end{figure}

Figure~\ref{fig:Qerror} shows errors, in \%, which come from the uncertainties 
of the energy levels.  
Most importantly, both methods A and B, see above, result in similar, relatively small errors 
(less than 0.004\% in the case of $Q_{\rm int}$ and less than 0.05\% in the case of $Q^{''}$). 
$C_{p}$ is the thermochemical quantity most sensitive to uncertainties of energy levels; 
therefore, the above analysis was repeated for $C_{p}$ 
(see the right panels of Figs.~\ref{fig:QconvT} and \ref{fig:QconvE} ).

\begin{figure}
\includegraphics[width=0.85\textwidth]{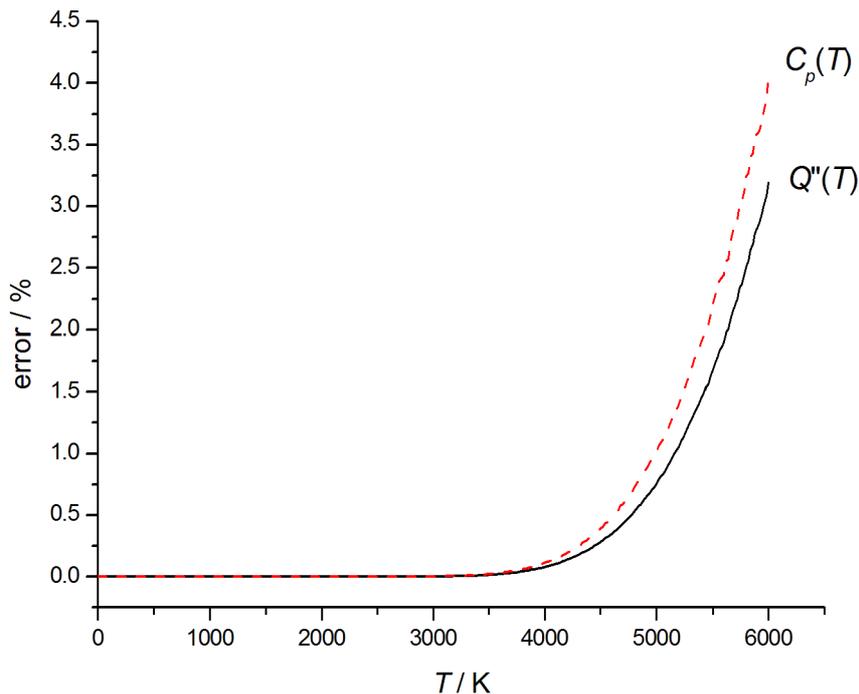}
\caption{Error due to the neglect of the unbound states 
during the determination of $Q''_{\rm int}(T)$  (solid line) and $C_{p}(T)$ (dashed line).}
\label{fig:Errorunbound}
\end{figure}

Figure~\ref{fig:Errorunbound} shows the error contribution of unbound states. 
It can be seen that (a) up to 4000 K the contribution is very close to zero; 
and (b) at 6000 K the contribution is 3.2\% to $Q''_{\rm int}(T)$ and 4.0\% to $C_{p}$.

\begin{figure}
\includegraphics[width=0.75\textwidth]{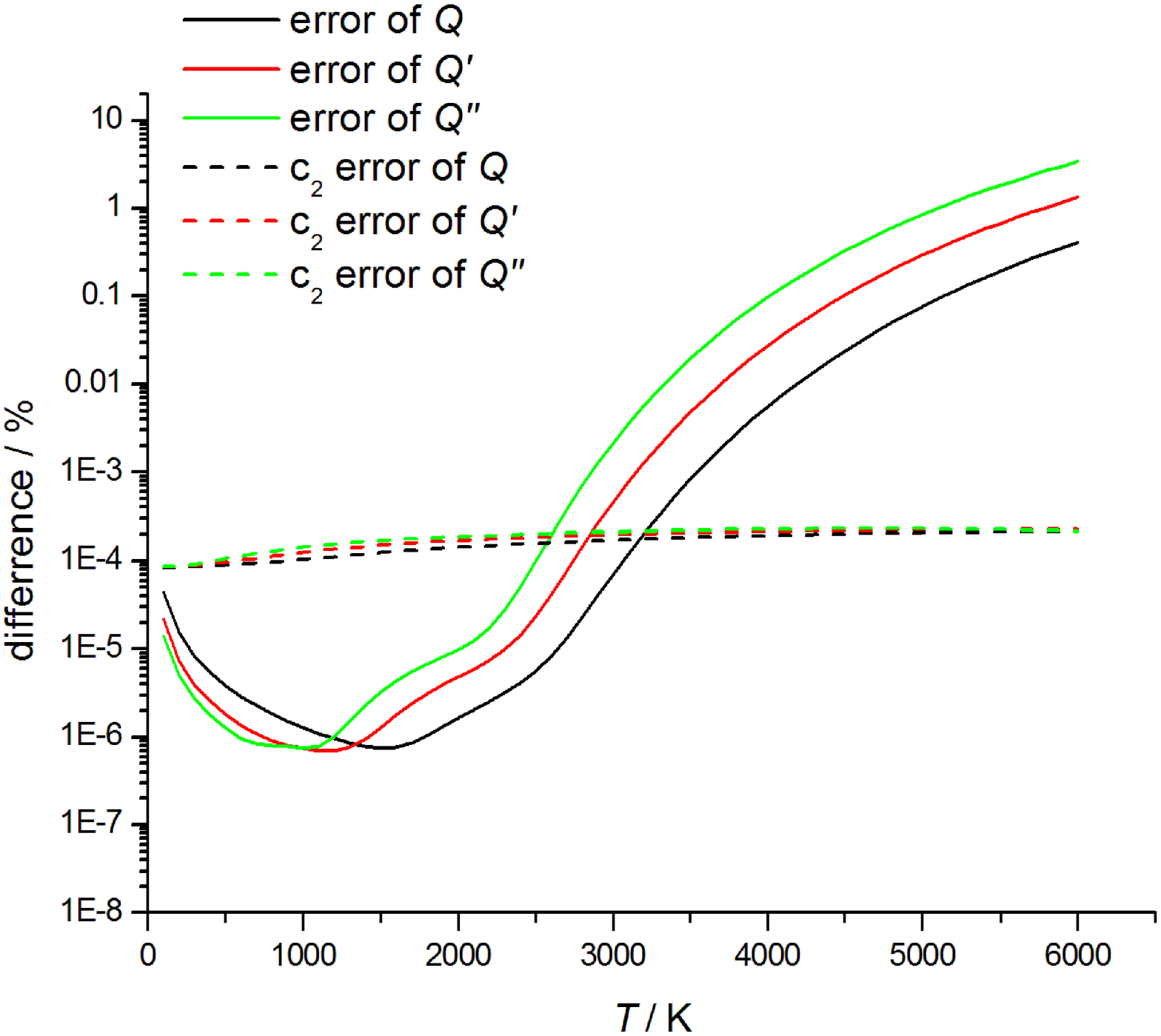}
\caption{Error contribution of energy levels (solid lines) and the error contribution of the uncertainty of second radiation constant (dashed lines).}
\label{fig:c2error}
\end{figure}

There is one more source of error which might influence the 
final uncertainty of the partition function: the uncertainties of the physical constants. 
This type of uncertainty is usually negligible, for example, in case of the 
heat capacity the uncertainty of the molar gas constant is two
orders of magnitude less than the other errors and 
since $R$ is a simple scale factor its uncertainty is negligible. 
However, in the case of $c_{2}$, the second radiation constant, 
which is the scale factor of energy levels and is inside the sum, 
the uncertainty of $c_{2}$ is not negligible. 
Figure~\ref{fig:c2error} shows the effect of an assumed error of $c_{2}$. 
It can be seen that below about 2500 K the uncertainty of $c_{2}$ 
determines the final uncertainty of the partition function. 
Above this temperature, the uncertainty contribution of unbound states dominates. 
The final uncertainty of the partition function is given by 
the four uncertainties just described.

\renewcommand{\baselinestretch}{0.765}
{\footnotesize
\begin{table*}[htbp]
\caption{\label{tab:QQprimeQdprime}The temperature-dependent internal partition functions of 
{\it ortho}- and {\it para}-H$_2$$^{16}$O, $Q_{\rm int}^{ortho}(T)$ and $Q_{\rm int}^{para}(T)$,
respectively, and their first two moments, $Q'$ and $Q''$.
The same data are also presented for the nuclear-spin-equilibrated quantity 
$Q_{\rm int}(T)$.
Numbers in parentheses are the approximate two standard 
deviation uncertainties in the last digits of the $Q_{\rm int}(T)$,
$Q'_{\rm int}(T)$, and $Q''_{\rm int}(T)$ data.}
					\begin{center}
          \resizebox{\columnwidth}{!}{%
          \begin{tabular}{rlllllllllll}
\hline \hline
		 $T$/K & $Q_{\rm int}^{para}(T)$ & $Q_{\rm int}^{\prime para}(T)$  & $Q_{\rm int}^{\prime\prime para}(T)$ & $Q_{\rm int}^{ortho}(T)$ & $Q_{\rm int}^{\prime ortho}(T)$  & $Q_{\rm int}^{\prime\prime ortho}(T)$ & $Q_{\rm int}(T)$ & $Q'_{\rm int}(T)$  & $Q''_{\rm int}(T)$  \\
		\hline
100 	&	8.78855	&12.79649&31.8905	&26.36458	&	38.40031	&	95.5838	&	35.15312(6)	&	51.19680(5)	&	127.4743(1)	\\
200 	&	24.3538	&36.0975&	90.3065	&	73.0613	&	108.2926	&	270.9194	&	97.4151(1)	&	144.3901(1)	&	361.2259(3)	\\
300 	&	44.5301	&66.5958&168.2010	&133.5904	&	199.7875	&	504.603	&	178.1206(2)	&	266.3833(2)	&	672.8040(6)	\\
400 	&	68.6423	&104.0492&268.974	&205.9269	&	312.1475	&	806.921	&	274.5692(3)	&	416.1966(4)	&	1075.895(1)	\\
500 	&	96.5825	&149.5104&399.179	&289.7475	&	448.5312	&	1197.536	&	386.3300(4)	&	598.0417(6)	&	1596.715(2)	\\
600 	&	128.5401&204.4202&565.317	&385.6204	&	613.2604	&	1695.952	&	514.1605(5)	&	817.6806(8)	&	2261.269(3)	\\
700 	&	164.8516&270.371&	774.746	&494.5548	&	811.114	  &	2324.238	&	659.4065(6)	&	1081.485(1)	&	3098.984(4)	\\
800 	&	205.945	&349.173&	1036.321&	617.835	&	1047.518	&	3108.963	&	823.7801(8)	&	1396.690(2)	&	4145.284(5)	\\
900  	&	252.325	&442.889&	1360.220&	756.976	&	1328.668	&	4080.66	&	1009.301(1)	&	1771.558(2)	&	5440.879(7)	\\
1000	&	304.568	&553.832&	1757.64	&	913.705	&	1661.495	&	5272.91	&	1218.273(1)	&	2215.327(3)	&	7030.54(1)	\\
1100	&	363.316	&684.527&	2240.64	&1089.948	&	2053.581	&	6721.91	&	1453.264(2)	&	2738.107(4)	&	8962.55(1)	\\
1200	&	429.272	&837.702&	2822.18	&1287.815	&	2513.107	&	8466.54	&	1717.087(2)	&	3350.809(5)	&	11288.73(2)	\\
1300	&	503.196	&1016.28&	3516.21	&1509.587	&	3048.84	  &	10548.61	&	2012.783(2)	&	4065.120(6)	&	14064.82(2)	\\
1400	&	585.903	&1223.381&4337.74	&1757.710	&	3670.141	&	13013.2	&	2343.613(3)	&	4893.522(7)	&	17350.94(4)	\\
1500	&	678.263	&1462.334&5303.00	&2034.789	&	4387.001	&	15908.99	&	2713.052(3)	&	5849.335(9)	&	21211.99(5)	\\
1600	&	781.198	&1736.69&	6429.52	&2343.593	&	5210.07	  &	19288.55	&	3124.790(4)	&	6946.76(1)	&	25718.07(8)	\\
1700	&	895.684	&2050.24&	7736.2	&2687.053	&	6150.72	  &	23208.6	&	3582.737(5)	&	8200.96(2)	&	30944.8(1)	\\
1800	&	1022.756&2407.02&	9243.4  &3068.268	&	7221.04	  &	27730.1	&	4091.024(6)	&	9628.06(3)	&	36973.5(2)	\\
1900	&	1163.504&2811.31&10973.0  &	3490.511&	8433.94	  &	32918.9	&	4654.015(7)	&	11245.25(4)	&	43891.8(3)	\\
2000	&	1319.078&3267.71&12948.5	&	3957.232&	9803.12	  &	38845.3	&	5276.309(9)	&	13070.83(5)	&	51793.8(4)	\\
2100	&	1490.69	&3781.06&	15195.0	&	4472.07	&	11343.16	&	45585	&	5962.76(1)	&	15124.22(7)	&	60780.0(6)	\\
2200	&	1679.61	&	4356.5&	17739.6	&	5038.84	&	13069.6	&	53218.7	&	6718.46(2)	&	17426.1(1)	&	70958.3(8)	\\
2300	&	1887.20	&	4999.6&	20611	&	5661.58	&	14998.7	&	61833	&	7548.78(2)	&	19998.3(1)	&	82444(1)	\\
2400	&	2114.84	&	5716.0&	23840	&	6344.52	&	17148.1	&	71519	&	8459.37(3)	&	22864.1(2)	&	95359(1)	\\
2500	&	2364.04	&	6512.0&	27459	&	7092.10	&	19536	&	82376	&	9456.14(4)	&	26048.0(2)	&	109835(2)	\\
2600	&	2636.33	&	7394.0&	31503	&	7908.99	&	22181.9	&	94508	&	10545.33(5)	&	29575.9(3)	&	126011(2)	\\
2700	&	2933.36	&	8368.9&	36009	&	8800.08	&	25106.6	&	108025	&	11733.44(6)	&	33475.4(4)	&	144034(3)	\\
2800	&	3256.83	&	9443.9&	41015	&	9770.49	&	28331.5	&	123045	&	13027.32(8)	&	37775.4(5)	&	164061(4)	\\
2900	&	3608.54	&10626.6&	46564	&	10825.59	&	31879.7	&	139693	&	14434.12(9)	&	42506.3(6)	&	186257(5)	\\
3000	&	3990.3	&11925.1&	52700	&	11971.0	&	35775.1	&	158099	&	15961.3(1)	&	47700.2(8)	&	210798(6)	\\
3100	&	4404.2	&13347.8&	59467	&	13212.6	&	40043.2	&	178401	&	17616.8(1)	&	53391(1)	&	237868(9)	\\
3200	&	4852.2	&	14904	&	66915	&	14556.5	&	44710	&	200744	&	19408.7(2)	&	59614(1)	&	267659(14)	\\
3300	&	5336.4	&	16602	&	75094	&	16009.2	&	49805	&	225282	&	21345.6(2)	&	66406(2)	&	300376(22)	\\
3400	&	5859.1	&	18452	&	84057	&	17577.3	&	55355	&	252171	&	23436.4(3)	&	73807(2)	&	336229(37)	\\
3500	&	6422.6	&	20464	&	93860	&	19267.8	&	61393	&	281578	&	25690.4(3)	&	81857(4)	&	375438(60)	\\
3600	&	7029.3	&	22649	&	104558	&	21087.9	&	67948	&	313673	&	28117.2(5)	&	90598(5)	&	418231(95)	\\
3700	&	7681.8	&	25018	&	116211	&	23045.2	&	75055	&	348633	&	30727.0(6)	&	100074(9)	&	464844(150)	\\
3800	&	8382.6	&	27583	&	128880	&	25147.6	&	82748	&	386639	&	33530.1(9)	&	110330(13)	&	515519(231)	\\
3900	&	9134	  &	30354	&	142626	&	27403	&	91061	&	427877	&	36538(1)	&	121415(20)	&	570503(347)	\\
4000	&	9940	  &	33344	&	157514	&	29820	&	100033	&	472541	&	39760(2)	&	133377(30)	&	630055(519)	\\
4100	&	10803	  &	36567	&	173605	&	32408	&	109700	&	520816	&	43210(3)	&	146266(45)	&	694421(751)	\\
4200	&	11725	  &	40034	&	190969	&	35175	&	120100	&	572907	&	46899(4)	&	160134(65)	&	763876(1081)	\\
4300	&	12710	  &	43758	&	209664	&	38130	&	131275	&	628991	&	50840(6)	&	175033(93)	&	838655(1507)	\\
4400	&	13761	  &	47755	&	229766	&	41284	&	143264	&	689298	&	55045(8)	&	191019(132)	&	919065(2103)	\\
4500	&	14882	  &	52036	&	251336	&	44645	&	156109	&	754007	&	59527(12)	&	208145(184)	&	1005342(2883)	\\
4600	&	16075	  &	56617	&	274434	&	48225	&	169851	&	823301	&	64300(17)	&	226468(253)	&	1097735(3874)	\\
4700	&	17345	  &	61512	&	299137	&	52034	&	184534	&	897411	&	69378(23)	&	246046(344)	&	1196548(5176)	\\
4800	&	18694	  &	66734	&	325519	&	56081	&	200202	&	976557	&	74775(31)	&	266936(462)	&	1302076(6882)	\\
4900	&	20126	  &	72299	&	353641	&	60379	&	216897	&	1060923	&	80505(42)	&	289196(613)	&	1414564(9055)	\\
5000	&	21646	  &	78222	&	383513	&	64938	&	234665	&	1150541	&	86584(57)	&	312886(806)	&	1534055(11566)	\\
5100	&	23257	  &	84516	&	415277	&	69769	&	253549	&	1245833	&	93026(75)	&	338066(1050)	&	1661110(14817)	\\
5200	&	24962	  &	91199	&	448970	&	74885	&	273596	&	1346915	&	99847(98)	&	364794(1354)	&	1795885(18823)	\\
5300	&	26765	  &	98283	&	484755	&	80296	&	294850	&	1454269	&107062(127)	&	393133(1731)	&	1939024(24100)	\\
5400	&	28672	  &	105786&	522468	&	86015	&	317357	&	1567409	&114687(164)	&	423142(2195)	&	2089877(29895)	\\
5500	&	30685	  &	113721&	562432	&	92054	&	341163	&	1687302	&122739(209)	&	454884(2761)	&	2249734(37402)	\\
5600	&	32809	  &	122105&	604338	&	98425	&	366314	&	1813022	&131234(265)	&	488418(3449)	&	2417360(45318)	\\
5700	&	35047	  &	130952&	648770	&	105141&	392856	&	1946321	&140188(333)	&	523808(4276)	&	2595091(55918)	\\
5800	&	37405	  &	140279&	695568	&	112214&	420837	&	2086714	&149619(416)	&	561115(5267)	&	2782281(68522)	\\
5900	&	39886	  &	150100&	744478	&	119657&	450301	&	2233449	&159543(515)	&	600401(6446)	&	2977927(82113)	\\
6000	&	42494	  &	160433&	796005	&	127483&	481299	&	2388030	&169977(635)	&	641731(7842)	&	3184035(98690)	\\
\hline \hline
          \end{tabular}}
        \end{center}
\end{table*}
}
\renewcommand{\baselinestretch}{0.925}

\begin{table*}[htbp]
\caption{\label{tab:ThermoChem}Thermochemical functions of 
nuclear-spin-equilibrated H$_2$$^{16}$O. 
Numbers in parentheses are the approximate two standard deviation uncertainties 
in the last digit of the quoted value.}
        \begin{center}
          \resizebox{\columnwidth}{!}{%
					\begin{tabular}{rccccccrrcrrc}
\hline \hline
\multicolumn{1}{c}{$T$/K}	& \multicolumn{3}{c}{$C_{p}(T)$ / J~K$^{-1}$~mol$^{-1} $} && \multicolumn{3}{c}{$S(T)$ / J~K$^{-1}$~mol$^{-1} $} && \multicolumn{3}{ c }{$H(T)$ / kJ~mol$^{-1} $} \\ \cline{2-4} \cline{6-8} \cline{10-12}
	
      &This work & Ruscic \cite{13Ruscic}& VT \cite{jt263} &&This work$^a$ & Ruscic \cite{13Ruscic}& VT \cite{jt263} &&This work & Ruscic \cite{13Ruscic}& VT \cite{jt263} \\ \hline  
100	&	33.30086(1)	&	33.301	&	33.301	&	&	152.38263(7)	&	152.387	&	152.384	&	&	3.28953(1)	&	3.290	&	3.289	\\
200	&	33.35053(1)	&	33.351	&	33.351	&	&	175.47984(7)	&	175.484	&	175.481	&	&	6.62199(1)	&	6.622	&	6.622	\\
300	&	33.59584(1)	&	33.596	&	33.596	&	&	189.03614(7)	&	189.040	&	189.038	&	&	9.96618(1)	&	9.966	&	9.966	\\
400	&	34.26208(1)	&	34.262	&	34.262	&	&	198.78271(8)	&	198.787	&	198.784	&	&	13.35574(1)	&	13.356	&	13.356	\\
500	&	35.22593(1)	&	35.226	&	35.226	&	&	206.52794(8)	&	206.532	&	206.530	&	&	16.82850(1)	&	16.829	&	16.829	\\
600	&	36.32471(1)	&	36.325	&	36.325	&	&	213.04616(9)	&	213.050	&	213.048	&	&	20.40529(1)	&	20.405	&	20.405	\\
700	&	37.49627(1)	&	37.496	&	--	    &	&	218.73282(9)	&	218.737	&	--~~~	  &	&	24.09582(1)	&	24.096	&	--~~~ \\	
800	&	38.72398(1)	&	38.724	&	38.728	&	&	223.8193(1)	  &	223.823	&	223.822	&	&	27.90642(1)	&	27.906	&	27.907	\\
900	&	39.99172(1)	&	39.99	&	--	&	&	228.4533(1)	&	228.457	&	--~~~	&	&	31.84196(1)	&	31.842	&	--~~~ \\	
1000&	41.27527(1)	&	41.269	&	41.287	&	&	232.7332(1)	&	232.737	&	232.737	&	&	35.90529(1)	&	35.905	&	35.907	\\
1100&	42.54776(1)	&	42.529	&	--	&	&	236.7271(1)	&	236.730	&	--~~~	&	&	40.09664(1)	&	40.095	&	--~~~ \\	
1200&	43.78553(2)	&	43.739	&	43.809	&	&	240.4826(1)	&	240.482	&	240.490	&	&	44.41368(1)	&	44.409	&	44.419	\\
1300&	44.97088(3)	&	44.872	&	--	&	&	244.0345(1)	&	244.029	&	--~~~	&	&	48.85199(1)	&	48.840	&	--~~~ \\	
1400&	46.09248(5)	&	45.909	&	46.124	&	&	247.4087(2)	&	247.393	&	247.420	&	&	53.40573(1)	&	53.380	&	53.417	\\
1500&	47.14441(7)	&	46.835	&	47.177	&	&	250.6250(2)	&	250.592	&	250.639	&	&	58.06816(1)	&	58.018	&	58.082	\\
1600&	48.1249(1)	&	--	&	48.157	&	&	253.6993(2)	&	--	&	253.715	&	&	62.83222(2)	&	--~~~	&	62.850	\\
1800&	49.8780(2)	&	--	&	49.904	&	&	259.4715(4)	&	--	&	259.491	&	&	72.63700(5)	&	--~~~	&	72.660	\\
2000&	51.3787(3)	&	--	&	51.394	&	&	264.8064(6)	&	--	&	264.828	&	&	82.7666(1)	&	--~~~	&	82.794	\\
2200&	52.6646(4)	&	--	&	52.668	&	&	269.7651(9)	&	--	&	269.788	&	&	93.1742(2)	&	--~~~	&	93.204	\\
2400&	53.7730(5)	&	--	&	53.766	&	&	274.396(1)	&	--	&	274.418	&	&	103.8206(3)	&	--~~~	&	103.850	\\
2600&	54.7373(6)	&	--	&	54.724	&	&	278.739(2)	&	--	&	278.761	&	&	114.6738(4)	&	--~~~	&	114.701	\\
2800&	55.5848(7)	&	--	&	55.571	&	&	282.827(2)	&	--	&	282.848	&	&	125.7077(5)	&	--~~~	&	125.732	\\
3000&	56.337(1)	&	--	&	56.326	&	&	286.689(3)	&	--	&	286.708	&	&	136.9013(7)	&	--~~~	&	136.923	\\
3200&57.008(4)&	--	&	57.005	&	&	290.346(4)	&	--	&	290.365	&	&	148.237(1)	&	--~~~	&	148.257	\\
3400&57.608(9)&	--	&	57.614	&	&	293.821(7)	&	--	&	293.840	&	&	159.700(2)	&	--~~~	&	159.720	\\
3600&	58.14(2)&	--	&	58.152	&	&	297.13(1)	&	--	&	297.149	&	&	171.276(5)	&	--~~~	&	171.298	\\
3800&	58.60(4)&	--	&	58.613	&	&	300.28(3)	&	--	&	300.305	&	&	182.950(9)	&	--~~~	&	182.976	\\
4000&	58.98(7)&	--	&	58.986	&	&	303.30(5)	&	--	&	303.322	&	&	194.71(2)	&	--~~~	&	194.737	\\
4200&	59.3(1)	&	--	&	59.259	&	&	306.19(9)	&	--	&	306.207	&	&	206.54(4)	&	--~~~	&	206.564	\\
4400&	59.5(2)	&	--	&	59.418	&	&	308.9(2)	&	--	&	308.968	&	&	218.41(7)	&	--~~~	&	218.433	\\
4600&	59.6(4)	&	--	&	59.451	&	&	314.1(4)	&	--	&	311.610	&	&	242.2(2)	&	--~~~	&	230.322	\\
4800&	59.6(5)	&	--	&	59.350	&	&	315.4(5)	&	--	&	314.139	&	&	248.2(2)	&	--~~~	&	242.205	\\
5000&	59.5(6)	&	--	&	59.111	&	&	316.6(6)	&	--	&	316.557	&	&	254.2(3)	&	--~~~	&	254.053	\\
5200&	59.3(8)	&	--	&	58.734	&	&	318.9(9)	&	--	&	318.868	&	&	266.0(4)	&	--~~~	&	265.840	\\
5400&	59(1)	  &	--	&	58.225	&	&	321(1)	&	--	&	321.076	&	&	277.9(6)	&	--~~~	&	277.538	\\
5600&	59(1)	  &	--	&	57.591	&	&	323(2)	&	--	&	323.182	&	&	289.7(9)	&	--~~~	&	289.122	\\
5800&	58(2)	  &	--	&	56.846	&	&	325(2)	&	--	&	325.191	&	&	301(1)	&	--~~~	&	300.567	\\
6000&	58(2)	  &	--	&	56.003	&	&	327(3)	&	--	&	327.104	&	&	313(2)	&	--~~~	&	311.854	\\

\hline \hline
\multicolumn{12}{l}{$^a$ The values reported in this column correspond to 
$S(T) -R \times {\rm ln}4$, to make the $S(T)$ results}\\
\multicolumn{12}{l}{of the present study approximately comparable to those of
Refs.~\citen{13Ruscic} and \citen{jt263}.}
          \end{tabular}}
        \end{center}
\end{table*}

\section{~~Results and Discussion}
The $Q$, $Q'$, and $Q''$ results of {\it ortho}- and {\it para}-H$_2$$^{16}$O, 
along with the nuclear-spin-equilibrated mixture, are presented in Table~2 in 100 K 
intervals up to 6000 K, starting at 100 K. 
Table~3 contains three thermochemical functions, $C_{p}(T)$, $S(T)$, and
$H(T)$ as a function of temperature, 
as well as their comparison with the best previous results 
obtained by Ruscic \cite{13Ruscic} and VT.\cite{jt263}
Note that only the traditional, nuclear-spin-equilibrated 
thermochemical quantities are compared with literature data in Table~3.
The complete set of results at 1 K increments is given in the 
Supplementary Material \cite{SupplMat} to this paper. 
Table~\ref{tab:Fit} lists coefficients of a least-squares fit to our computed 
partition function, using the traditional form of \cite{jt263}
\begin{eqnarray}
{\rm \ln}Q_{\rm int}=\sum\limits_{i=0}^{6} a_i({\rm ln}T)^i.
\label{Qint}
\end{eqnarray}
In order to get the best reproduction of the directly computed values, 
the fit had to be performed in two separate temperature ranges.
The first range is $0-200$ K, the other is $201-6000$ K. 
These fits can reproduce, in both regions, 
the values of ln$Q_{\rm int}$ reasonably accurately, within about 0.1\%.
Nevertheless, as emphasized, for example, by Fischer et al.,\cite{03FiGaGo.partfunc} 
it is preferable these days to interpolate the tabulated thermochemical data presented
in a fine grid rather than to use low-order polynomical expansions, and
even a 25~K tabulation is sufficient for most partition functions.

\begin{table*}[ht]
\caption{Coefficients of the fit, see Eq. (17), to the nuclear-spin-equilibrated 
internal partition function of H$_2$$^{16}$O.}\label{tab:Fit}
        \begin{center}
          \begin{tabular}{crrr}
\hline \hline
		 Coefficient  & 0 -- 200 K~~ && 201 -- 6000 K \\
		\hline
$a_0$  &   0.0000414145 &&   86.9112357472 \\
$a_1$  &   5.2668268683 &&  --60.7285954830 \\
$a_2$  &  --8.4438709824&&   15.4447694151 \\
$a_3$  &   4.9777150504 &&  --1.3899526096 \\
$a_4$  &  --1.3449867842&&  --0.0424069070 \\
$a_5$  &   0.1743063797 &&    0.0143520410 \\
$a_6$  &  --0.0087936229&&  --0.0006287480 \\
\hline \hline
          \end{tabular}
        \end{center}
\end{table*}

\subsection{~~~~The partition function}
What the dimensionless partition function tells us is basically 
the ratio of the total 
number of ``particles'' to the number of ``particles'' in the ground state.
Thus, the partition function provides a greatly simplified measure 
of how the particles are partitioned among the available energy levels.
The magnitude of the partition function depends upon the magnitude of the 
fractional populations, and the latter depend both on the relative energy 
of the state and the chosen temperature.
The largest value of the partition function can be very large 
but not infinite if the system contains a finite number of ``particles''.
As seen in Table 2, $Q_{\rm int}(T)$ of H$_2$$^{16}$O becomes large as the 
temperature increases,
reaching about (1, 1200, 5300, 16~000, 40~000, 87~000, 170~000) at temperatures of 
(0, 1000, 2000, 3000, 4000, 5000, 6000) K.

\begin{figure}
\includegraphics[width=0.75\textwidth]{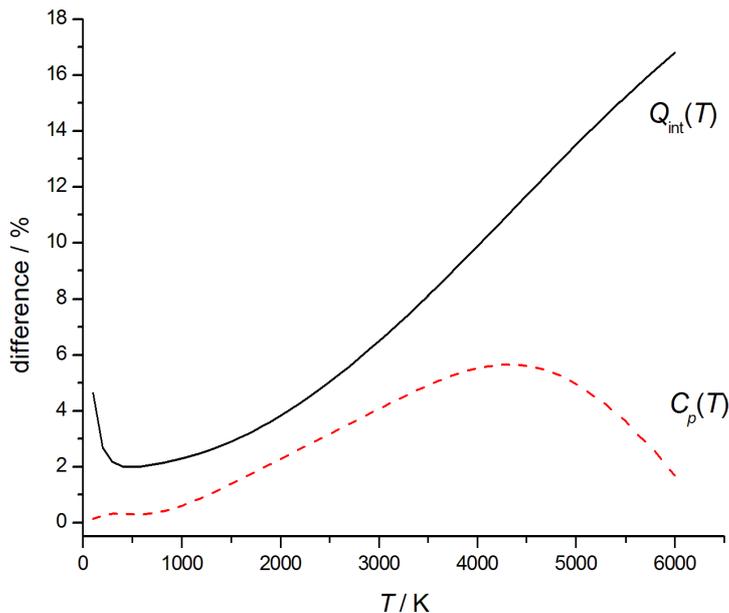}
\caption{Percentage difference between the ``exact'' values of $Q_{\rm int}(T)$
and $C_{p}(T)$
and those corresponding to the ``analytical'' rigid-rotor--harmonic-oscillator 
(RRHO) approximation.
The apparent increase in the differences below 350~K 
for $Q_{\rm int}(T)$ is due to the failure of 
approximating the direct sum with an integral when the density of states is low.}
\label{fig:diffrrho}
\end{figure}

\subsection{~~~~Comparison with previous results}
The simplest way to approximate the partition function is 
through the application of the rigid-rotor and harmonic-oscillator (RRHO) model.
For the RRHO model an analytical formula is known for generating the partition function. 
Using experimental spectroscopic constants 
($A$ = 835 839.9 MHz, $B$ = 435 354.5 MHz, and $C$ = 278 133.3 MHz from Ref.~\citen{jt355};
$\nu_{1}$ = 3657.053251 \cm, $\nu_{2}$ = 1594.746292 \cm, and $\nu_{3}$ = 3755.928548 \cm\ 
from Ref.~\citen{jt539}), the temperature-dependent internal partition function 
can easily be computed. 
Figure~\ref{fig:diffrrho} shows the differences for $Q_{\rm int}(T)$
and $C_p(T)$ between the ``exact'' values of this study
and those of the ``analytical'' RRHO partition function.
Although the difference between the exact and the RRHO values can be significant
for $Q_{\rm int}(T)$, especially at higher temperatures, 
considering the extreme simplicity of this model the agreement observed
is quite pleasing for this semirigid molecule.
Note also that most of the difference between the discrete, ``exact'' results and 
the continuous, ``analytical'' RRHO results at the lowest temperatures, 
below about 350 K, is due to failure of the integration approximation.
The differences would tend toward zero if the experimental spectroscopic constants
reported were used to generate rovibrational energy levels and these were used,
via direct summation, for the computation of $Q_{\rm int}(T)$.
Note that, in a relative sense, the RRHO approximation works seemingly considerably better
for $C_p(T)$ than for $Q_{\rm int}(T)$.

\begin{figure}   
\includegraphics[width=0.485\textwidth]{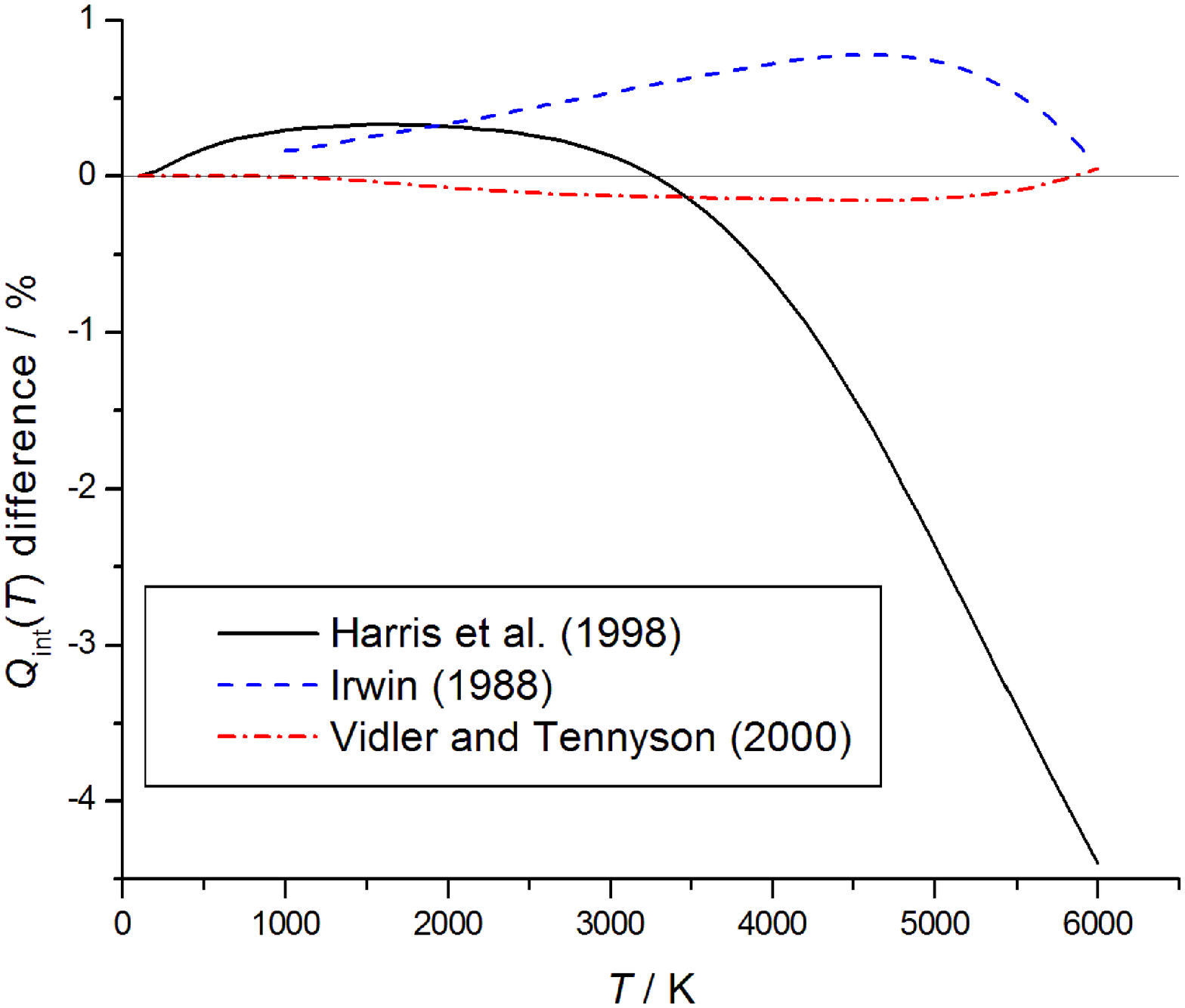}
\includegraphics[width=0.485\textwidth]{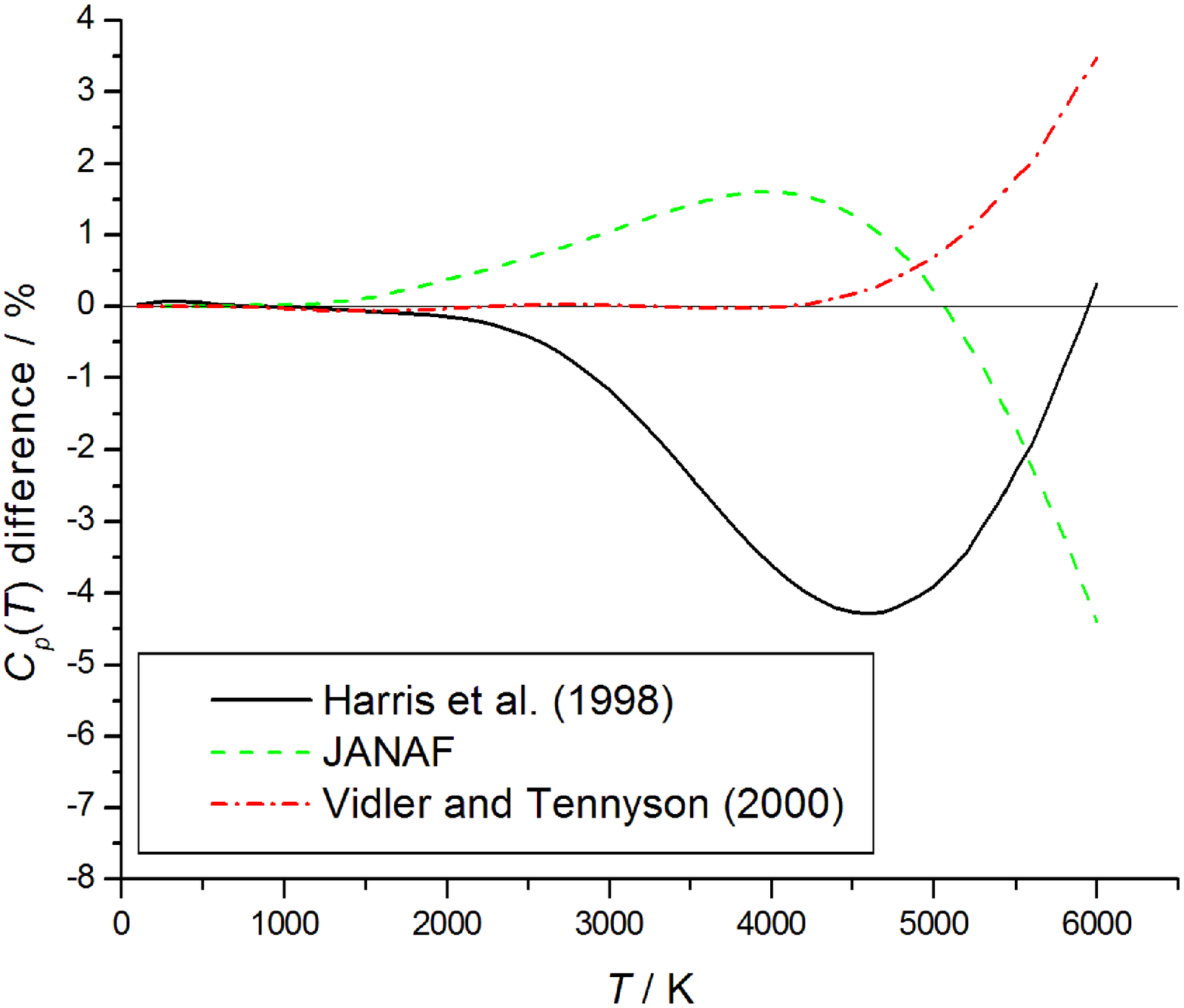}
\caption{Left panel: Comparison of the present $Q_{\rm int}$($T$) values with those of
Harris {\it et al.},$^{29}$ Irwin,$^{66}$ and Vidler and Tennyson.$^{30}$
Right panel: Comparison of the present $C_{p}$($T$) values with those of
Harris {\it et al.},$^{29}$ JANAF,$^{15}$ and Vidler and Tennyson.$^{30}$}
\label{fig:comp}
\end{figure}

The left panel of Fig.~\ref{fig:comp} shows 
the comparison of our internal partition function with other 
high-temperature values by Harris {\it et al.},\cite{jt228} 
Irwin,\cite{88Irwin} and VT.\cite{jt263}
The agreement with the VT results is especially pleasing.
The right panel of Fig.~\ref{fig:comp} compares our $C_{p}$($T$) values with 
those of Harris {\it et al.},\cite{jt358} JANAF,\cite{janaf4} and VT.\cite{jt263}
As expected, the deviations here are slightly larger, 
but VT works very well below 4500 K.

\subsection{~~~~NASA polynomials}
The tabular form of thermochemical data used to be 
not very convenient for computerized applications.  
Thus, more than four decades ago Gordon and McBride \cite{92McGo} 
suggested a set of low-order polynomials 
providing a convenient set of fit functions known as the older 7-constant 
and the newer 9-constant NASA polynomials.  
As Ruscic {\it et al.} \cite{05IUPAC} emphasized, 
(a) the 9-constant NASA polynomial reproduces the underlying data about two orders of magnitude 
better than the 7-constant NASA polynomial, and 
(b) the thermochemical properties can be calculated in general with confidence 
in the fourth and fifth digit in the range of $150-3000$~K.
Nevertheless, the accuracy of even the 9-constant NASA polynomial is clearly insufficient 
when the data of the present study are considered.
Therefore, thermochemical quantities determined in this study are provided in the
Supplementary Material \cite{SupplMat} at 1~K intervals.
For those who need highly accurate thermochemical data, it is recommended to
adapt the tabulated functions.

\subsection{~~~~CODATA}
The outstanding accuracy of the experimental (MARVEL) energy levels employed in this study
means that all thermochemical quantities computed, especially at lower temperatures,
have exceedingly high accuracy (the list of MARVEL rovibrational energies is complete
up to 7500 \cm).
One such quantity is $H^{\rm o}$(298.15~K)--$H^{\rm o}$(0~K), 
the standard molar enthalpy increment (standard integrated heat capacity) of H$_2$$^{16}$O
between 298.15 and 0 K ($H^{\rm o}$(0~K) = 0.0 J~mol$^{-1}$).
This value is given for water in the official CODATA
compilation \cite{89CoWaMe} as $9.905 \pm 0.005$ kJ~mol$^{-1}$.
Naturally, the value we compute, $9.90404 \pm 0.00001$ kJ~mol$^{-1}$, 
falls within the limits of the old value, but it is more accurate 
by about two orders of magnitude.
The newly determined value is insensitive to any reasonable change in
the energy levels; the value is completely determined by energy levels lower than
about 5000 \cm, and, notably, even the present first-principles PoKaZaTeL energy levels 
yield the same value though with higher uncertainty.
While the present suggested change in the standard molar enthalpy increment 
of H$_2$$^{16}$O is more or less inconsequential for most of thermochemistry, as
enthalpies of formation cannot be determined with this exceedingly small uncertainty,
it nevertheless exemplifies the fact that it is more and more realistic to use high-resolution 
spectroscopic data to directly calculate thermodynamic quantities with minuscule uncertainties.

\newpage
\begin{figure}
\includegraphics[width=0.68\textwidth]{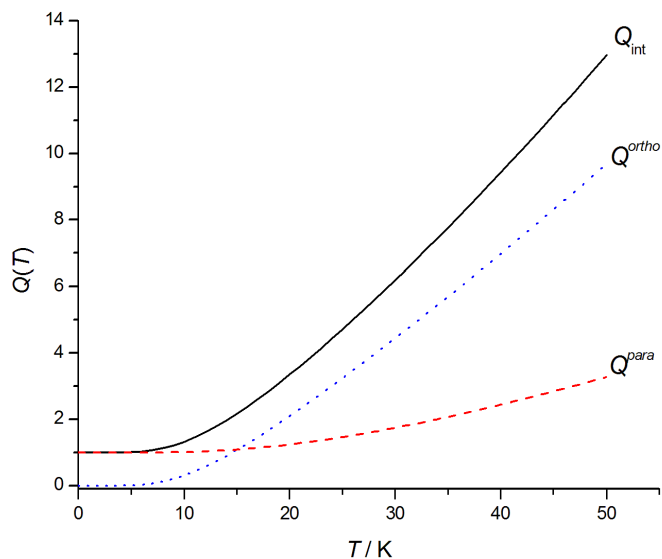}
\caption{The \textit{ortho}-H$_2$$^{16}$O (dotted, blue curve), 
the \textit{para}-H$_2$$^{16}$O (dashed, red curve), 
and the nuclear-spin-equilibrated H$_2$$^{16}$O  (full, black curve) 
partition functions at low temperatures, below 50~K.}
\label{fig:Qop}
\end{figure}

\subsection{~~~~The low-temperature limit}
Standard thermochemical textbooks and standard thermochemical tables 
found in various compendia \cite{89GuVeAl,janaf4}
venture very rarely below 100 K.
The reason is that there are special considerations about partition functions 
as well as thermochemical functions at the lowest temperatures, 
(well) below 100 K, due to the effect of nuclear spin statistics.
It is only for higher temperatures that the \textit{ortho} and \textit{para} spin isomers 
of water are equilibrated, while in thermochemistry one always assumes an equilibrated
mixture.
In fact, the effect of nuclear spins can be investigated on effective structural parameters, 
as has been done for H$_2$O \cite{09CzMaCs} and NH$_3$.\cite{12SzFaCzMa}
Due to the distinct rovibrational states, the \textit{ortho} and \textit{para} species
have slightly different effective structures and different thermochemical functions.
The two nuclear-spin isomers can be in equilibrium (note again that this is what we 
always assume in thermochemistry), or, if their interconversion is kinetically hindered,
they exist as a mixture corresponding to distinct nuclear spin
temperatures.\cite{jt330,jt349}
The same phenomenon is well known and has been studied \cite{27Dennison,29BoHa}
at the dawn of quantum mechanics for the H$_2$ molecule.

Figure~\ref{fig:Qop} shows the values of the internal partition functions
of the \textit{ortho} and \textit{para} spin isomers of H$_2$$^{16}$O below 50~K.
The different low-temperature behavior of 
$Q^{ortho}_{\rm int}$ and $Q^{para}_{\rm int}$ is
evident from the figure.
As mentioned, it is only the nuclear-spin-equilibrated $Q_{\rm int}$($T$) which is part of 
traditional thermochemistry.
The reader should also be warned that one should not mix thermochemical data 
adhering to different definitions, in this case the convention used to
represent whether nuclear spin effects are considered or not.

\begin{figure}
\includegraphics[width=0.75\textwidth]{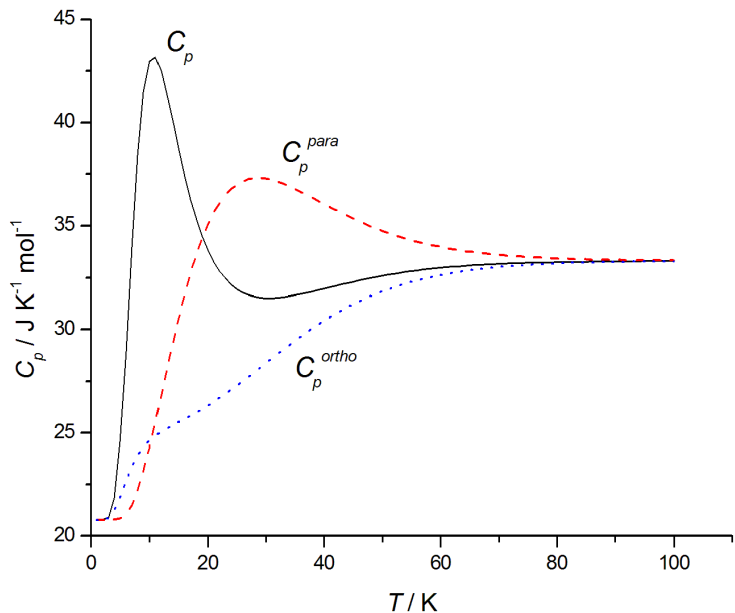}
\caption{
The \textit{ortho}-H$_2$$^{16}$O (dotted, blue curve), 
the \textit{para}-H$_2$$^{16}$O (dashed, red curve), 
and the nuclear-spin-equilibrated H$_2$$^{16}$O (full, black curve) 
isobaric heat capacities at low temperatures, below 100~K.}
\label{fig:Cpop}
\end{figure}

Figure \ref{fig:Cpop} shows the 
isobaric heat capacity of the \textit{ortho} and \textit{para} 
spin isomers as a function of temperature, as well as that of the 
equilibrium mixture.
It is seen that up to 80 K the two water isomers possess rather different
curves, but above 100 K the two curves become basically the same.

It must also be noted that the ortho-to-para (OPR) ratio 
is a useful diagnostic tool in astrochemistry.\cite{15FuAiHiHa}
The drastically different isobaric heat capacity of \textit{ortho}- and
\textit{para}-H$_2$$^{16}$O between 10~K and 60~K computed here with high accuracy
may have important consequences for certain applications.

\section{~~Summary and Conclusions}
The ideal-gas internal partition functions determined in this study for
\textit{ortho}-H$_2$O, \textit{para}-H$_2$O, and their nuclear-spin-equilibrated mixture, 
in the temperature range of 0--6000~K, are the most accurate ones produced to date.
The partition functions as well as the subsequently determined 
thermochemical functions, including 
the standardized enthalpy, the entropy, and the isobaric heat capacity, 
have their own temperature-dependent uncertainties.
All these thermochemical quantities
are listed in 100~K increments in the main text and in 1 K increments in the 
Supplementary Material;\cite{SupplMat} the latter should support several modeling applications.
The accuracy of the present data is due to the following characteristics of this study:

(1) The internal partition function $Q_{\rm int}$($T$) and its first two moments
are determined via the explicit summation technique; thus, their determination
involves no modeling assumptions beyond the bare basics, distinguishing this study
from almost all previous efforts.

(2) A large number of highly accurate, experimental rovibrational energy levels
determined previously \cite{jt539} are utilized; 
the list of experimental levels is complete up to 7500 \cm,
significantly lowering the uncertainty of $Q_{\rm int}$($T$) below about 1000 K.

(3) At higher temperatures, between about 1000 K and 3000~K,
the completeness of the energy level set
determines the true accuracy of the thermochemical quantities determined.
We utilized the complete set of bound rovibrational energy levels of H$_2$$^{16}$O
obtained from a first-principles variational nuclear motion computation involving an
exact kinetic energy operator and a highly accurate empirical PES.\cite{jtpoz}
Altogether close to one million bound energy levels are utilized in this study.

(4) In order to ensure accuracy between 3000 and 6000~K, the contribution due to 
unbound states is considered via a simple model computation.
Our test computations show that for H$_2$$^{16}$O
the contribution from the excited electronic states can be safely neglected.

Although in this study highly accurate thermochemical functions have been obtained 
for \textit{ortho}- and \textit{para}-H$_2$$^{16}$O, it is not yet common to include
nuclear-spin statistical factors in the computation of partition functions and the related
thermochemical functions.
Thus, these data should be used with caution in chemical reactions 
where nuclear spin effects are neglected for the other species involved.
Nevertheless, it is expected that such data, especially important at low temperatures,
will become available for a growing number of chemical species.

The accuracy of the reproduction of the present data with the 7- and 9-constant NASA
polynomials is orders of magnitude worse than the internal accuracy of our results.
Thus, it is recommended to use the 1~K list of computed values in all applications
requiring high accuracy.

It is recommended that the new, exceedingly accurate value of the
standard molar enthalpy increment (integrated heat capacity) of H$_2$$^{16}$O,
$H^{\rm o}$(298.15~K) = $9.90404(1)$ kJ~mol$^{-1}$, should replace the
value advocated in the CODATA compilation,\cite{89CoWaMe} $9.905 \pm 0.005$ kJ~mol$^{-1}$.

Finally, we note that the present procedure and data serve well the mission of IAPWS 
to determine accurate ideal- and real-gas data for water.
For this task we need similarly high-quality data for all water isotopologues
present in Vienna Standard Mean Ocean Water 
(VSMOW),\cite{78Gonfi,VSMOW,VSMOW2} providing the isotopic 
composition \cite{VSMOW} of so-called ``ordinary water substance.''
Work in this direction is in progress.

\begin{acknowledgments}
The authors are grateful to the COST action 
``Molecules in Motion'' (MOLIM, CM1405) for support.
AGC thanks the NKFIH (grant number NK83583 and K119658) 
for supporting the work performed in Hungary.
JH acknowledges support provided by the Czech Science
Foundation (grant no. 16-02647S).
The authors are indebted to Professors Roberto Marquardt 
and Branko Ruscic for their useful comments on certain aspects of this study.
\end{acknowledgments}


%
\end{document}